\documentclass{article}

\newcommand{\E}{{\mathbb{E}}}
\newcommand{\VAR}{\mathrm{Var}}
\newcommand{\COV}{\mathrm{Cov}}

\newcommand{\R}{{\mathbb{R}}}
\newcommand{\eps}{{\varepsilon}}

\newcommand{\X}{{\bm{X}}}
\newcommand{\F}{{\bm{F}}}
\newcommand{\Am}{{\text{Am}}}
\newcommand{\Gm}{{\text{Gm}}}

\newcommand{\Z}{{\bm{Z}}}
\newcommand{\W}{{\bm{W}}}
\newcommand{\eigunit}{{\lambda_{\text{unit}}}}
\newcommand{\Junit}{{J_{\text{unit}}}}

\newcommand{\Xbind}{{X_{\text{bind}}}}
\newcommand{\co}[2]{c_{#1}^{(#2)}}

\newcommand{\PIAR}[2]{\text{PI}_{#1}\text{AR}(#2)}

\newtheorem{definition}{Definition}[section]

\usepackage{arxiv}
\usepackage[utf8]{inputenc} 
\usepackage[T1]{fontenc}    
\usepackage[colorlinks=true]{hyperref}       
\usepackage{url}            
\usepackage{booktabs}       
\usepackage{amsfonts}       
\usepackage{nicefrac}       
\usepackage{microtype}      
\usepackage{lipsum}		
\usepackage{amsmath}
\usepackage{graphicx}
\usepackage{natbib}
\usepackage{float}
\usepackage{bm}
\usepackage{subcaption}
\usepackage{tabularx}
\usepackage{threeparttable}
\bibpunct{(}{)}{,}{a}{}{;}
\setcitestyle{round,aysep={,},yysep={;}}
\usepackage{doi}

\title{A Multi-Companion Method to Periodically Integrated Autoregressive Models}

\date{September 15, 2023}

\author{ \href{https://orcid.org/0000-0003-2066-4050}{\includegraphics[scale=0.06]{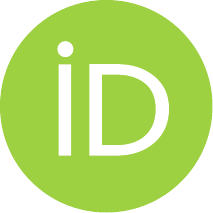}\hspace{1mm}Yueyun Zhu}
	\\
	Department of Mathematics\\
	The University of Manchester\\
	Manchester, M13 9PY \\
	\texttt{yueyun.zhu@manchester.ac.uk} \\
	\And
	\href{https://orcid.org/0000-0003-2839-346X}{\includegraphics[scale=0.06]{orcid.pdf}\hspace{1mm}Georgi N. Boshnakov} \\
	Department of Mathematics\\
	The University of Manchester\\
	Manchester, M13 9PY \\
	\texttt{georgi.boshnakov@manchester.ac.uk} \\
}

\renewcommand{\shorttitle}{Multi-Companion method for PIAR models}

\begin{document}
\maketitle

\begin{abstract}
	There has been an enormous interest in analysing and modelling periodic time series. The research on periodically integrated autoregressive (PIAR) models which capture the periodic structure and the presence of unit roots is widely applied in environmental, financial and energy areas. In this paper, we propose a multi-companion method which uses the eigen information of the multi-companion matrix in the multi-companion representation of PIAR models. The method enables the estimation and forecasting of PIAR models with a single, two and multiple unit roots. We show that the parameters of PIAR models can be represented in terms of the eigen information of the multi-companion matrix. Consequently, the estimation can be conducted using the eigen information, rather than directly estimating the parameters of PIAR models. A Monte Carlo experiment and an application are provided to illustrate the robustness and effectiveness of the multi-companion method.
\end{abstract}


\section{Introduction}
\label{Introduction}

The presence of strong periodicity and seasonal variations in financial, environmental and energy time series has inspired a critical area of research in periodic time series analysis. The analysis of periodic time series can be traced back as early as \cite{hannan1955test} and \cite{gladyshev1961periodically}, where the definition of periodic correlation and basic properties of periodic correlated series are given. Two books, see \cite{franses1996periodicity} and \cite{franses2004periodic}, give a comprehensive introduction to quarterly periodic time series models, including model representations, model selection, parameter estimation and forecasting. 

Periodic autoregressive (PAR) models have gained significant attention in recent years due to their ability to capture the periodic structure of the time series. The PAR models extend the conventional Autoregressive (AR) models by allowing the autoregressive parameters to change with seasons, and they are applied to analyse periodically stationary series. The early reference see \cite{pagano1978periodic} introduced a periodic Yule-Walker method to estimate periodic autoregressive parameters, and the continuous work done by \cite{troutman1979some} derived several key properties of autocovariance and its asymptotic properties of PAR models. Other estimation methods, such as maximum likelihood estimation \citep[see][]{vecchia1985periodic} and weighted least squares \citep[see][]{basawa2001large}, are also introduced to estimate PAR models. Later on, the topic of parsimonious PAR models has become popular because they use a minimal number of parameters to effectively capture the periodic behaviour of time series. The idea of parsimonious PAR models was initially proposed by \cite{jones1967time}, and the later references can be found in \cite{lund2006parsimonious}, \cite{anderson2007fourier}, \cite{tesfaye2011asymptotic} and \cite{battaglia2020parsimonious}.

Despite the popularity of PAR models, their estimation and prediction can be challenging when the time series is periodically non-stationary. Therefore, the periodically integrated autoregressive (PIAR) models are developed to deal with the periodic non-stationarity, incorporating with the concept of periodic integration and the presence of unit roots in the series.

Before delving into periodic integration, we first illustrate the (ordinary) integration for non-periodic cases. For non-periodic cases, the concept of integration is introduced to handle the presence of unit roots in the non-stationary series. A time series is said to be integrated of order $b$, denoted as $\text{I}(b)$, if it can be transformed into a stationary series by taking the $b$-th difference, while the first $(b-1)$ differences are non-stationary. Without loss of generality, we have the notation $\text{I}(0)$ to describe the stationary series. Based on the concept of integration, the autoregressive integrated moving average (ARIMA) models are then developed to link the autoregression, moving average and integration together to analyse non-periodic time series that exhibit non-stationarity. 

Similar to the non-periodic cases, the concept of periodic integration was introduced to handle the presence of unit roots in periodically non-stationary series. One of the earliest references is \cite{osborn1988seasonality}, who discussed the case when there is a single unit root in the periodic series and gave the definition of periodic integration of order one. The later work is followed by \cite{boswijk1995testing} who proposed three tests for checking the quarterly PIAR models with a single unit root. \cite{boswijk1996unit} proposed a class of likelihood ratio tests for a single unit root in PIAR models and derived their asymptotic distributions under the null hypothesis that there is a single unit root. \cite{boswijk1997multiple} extended the previous studies by proposing a new test which can be employed to check quarterly PIAR models with multiple unit roots. The model selection and forecasting issues of PIAR models can be found in \cite{franses1996periodic}. 

In this paper, we propose an innovative method, the multi-companion method, which is based on
the eigen information (eigenvectors and eigenvalues) of the multi-companion matrix in the
multi-companion representation of PIAR models. The multi-companion matrix can be viewed as a
generalization of the companion matrix and it was firstly introduced by
\cite{boshnakov2002multi}. \cite{boshnakov2002multi} derived several key properties of the
multi-companion matrices, which can be further applied to time series analysis. Due to these
special properties, the multi-companion matrices have been a topic of interest in recent
years. For instance, \cite{boshnakov2009generation} used the eigen information of the
multi-companion matrices to generate periodic autoregressive series. Also,
\cite{boshnakov2009generation} listed the algorithms for generating multi-companion matrices
by using their eigen information, and these algorithms prove to be highly beneficial in
simulation studies. An R \citep{R} implementation of multi-companion matrices, including
spectral parametrisation, is provided by \citet{RpackageMcompanion}. Further functionality,
specific for periodic models, is included in package pcts \citep{RpackagePCTS}.

It is worthwhile to mention that the previous studies, such as \cite{boswijk1996unit} and \cite{boswijk1997multiple}, mainly focused on quarterly PIAR models, and their methods may become inefficient when extending the quarterly period to general cases. In contrast, our multi-companion method proposed in this paper provides a more flexible and efficient way to analyse PIAR models with general periods. 

The paper is organized as follows. Section \ref{Models} reviews three different model representations for both PAR and PIAR models. Section \ref{Multi-companion method} proposes the multi-companion method, which decomposes the multi-companion matrix of the multi-companion representation into its Jordan canonical form, and the roles of similarity and Jordan matrices are illustrated respectively. Section \ref{Multi-companion method applied to PIAR model} applies the multi-companion method to analyse PIAR models with a single, two and multiple unit roots, and proposes an estimation method for PIAR models. In Section \ref{Monte Carlo Analysis}, Monte Carlo simulations are provided to verify the estimation method introduced in Section \ref{Multi-companion method applied to PIAR model}. Section \ref{Application} gives an application of PIAR models to forecast future values of U.S. monthly electricity end use.

It is useful to introduce some notation before going into detail. We use $d$ for denoting the period of the time series, $s$ for denoting the seasons such that $s \in \left\{1,\dots,d\right\}$. Any time $t$ can be represented equivalently by the year $T$ and the season $s$, such that $[T,s] \equiv t \equiv (T-1)d+s$, and therefore, we use the notation $[T,s]$ to refer time $t$ at year $T$ and season $s$ \citep[see][]{boshnakov2009generation}. The notation PAR($p$) is used to describe a periodic autoregression of order $p$. The notation $\PIAR{b}{p}$ is used to describe a periodically integrated autoregressive model with periodic integration order $b$ and periodic autoregression order $p$. Sometimes we may omit the periodic integration order $b$ to write as PIAR($p$) when the periodic integration order is unknown.


\section{Models}
\label{Models}

Let $\left\{X_t, t=1,2,\dots\right\}$ be a periodic time series with period
$d$. \cite{gladyshev1961periodically} defined that a process $\left\{X_t\right\}$ is said to
be periodically correlated (periodically stationary) with period $d$ if $\E[X_t]=\E[X_{t+d}]$
and $\COV(X_{\tau+d}, X_{t+d})=\COV(X_\tau,X_t)$ for all integer $\tau$ and
$t$. \cite{jones1967time} is probably the first study of periodic autoregressive models
(PAR). \cite{pagano1978periodic} obtained asymptotic properties of periodic Yule-Walker
estimators for PAR models.

\cite{osborn1988seasonality} proposed a concept of periodic integration for the case when the
series exhibits stochastic trends and therefore is no longer periodically
stationary. Periodically integrated autoregressive models for such periodically integrated
series have been studied by \cite{boswijk1996unit}, \cite{franses1996periodicity}, and
\cite{franses2004periodic}.

In this section, we will review various representations for both PAR and PIAR models. Importantly, a multi-companion representation will be given in the end of this section which is essential for further analysis.

\subsection{Univariate representation}
\label{Univariate representation}

We consider univariate periodic time series $\left\{X_t\right\}$ with a period of $d$ that can be transformed to white noise using a periodic filter:
\begin{align}
  \label{PARp Rep}
  X_t - \sum_{i=1}^{p}\phi_{i,s}X_{t-i} &= \eps_t
  , \qquad{}
  t = 1,2,\dots,
\end{align}
where $p$ is the periodic autoregressive order; $s=s(t) \in \left\{1,2,\dots,d\right\}$ is a function of time $t$ which returns the corresponding season index at time $t$, such that $s=d$ when $t \bmod d=0$, and otherwise, $s=t \bmod d$; $\left\{\phi_{i,s}, i=1,\dots,p\right\}$ are seasonally varying parameters with $d$-periodic such that $\phi_{i,s}=\phi_{i,s+d}$ for $s=1,\dots,d$; $\eps_t$ is a periodic white noise process with $\E[\eps_t]=0$, $\E[\eps_t\eps_{\tau}]=0$ when $t \neq \tau$, and $\VAR(\eps_t)=\sigma_t^2=\sigma_{t+d}^2$. The last property of the periodic white noise indicates that the variance of periodic white noise is $d$-periodic and it is sufficient to consider $\sigma_s^2$ for $s=1,\dots,d$ only. The notation $\eps_t \sim \text{PWN}(0,\sigma_s^2)$ is used to describe the periodic white noise with
variance $\sigma_s^2$, see \cite{boshnakov1996recursive} for details.

The left-hand side of Eq (\ref{PARp Rep}) represents a filter operation which is fully
described by the coefficients $\left\{\phi_{i,s}, i=1,\dots,p\right\}_{s=1}^{d}$.  Let
$\phi_{p,s}(z)=1-\phi_{1,s}z-\dots-\phi_{p,s}z^{p}$ be the polynomial associated with the
coefficients for the $s$th season, $s=1,\dots,d$. The set of polynomials
$\left\{\phi_{p,s}(z)\right\}_{s=1}^{d}$ can be used as an alternative way to specify the
filter.

The periodic filter of Eq (\ref{PARp Rep}) extends the conventional filter
$\phi_p(L)=1-\phi_1L-\dots-\phi_pL^p$ to allow the parameters changing with seasons. Unlike
the non-periodic filters, commutativity does not hold, in general, for periodic filters. To
demonstrate this, consider two periodic filters $\alpha_{1,s}(L)=1-\alpha_{1,s}L$ and
$\beta_{1,s}(L)=1-\beta_{1,s}L$. Let $\{\alpha_{1,s}(L)\beta_{1,s}(L)\}_{s=1}^{d}$ be the
filter corresponding to first applying $\{\beta_{1,s}(L)\}_{s=1}^{d}$ to the series, then
$\{\alpha_{1,s}(L)\}_{s=1}^{d}$ to the filtered series and similarly
$\{\beta_{1,s}(L)\alpha_{1,s}(L)\}_{s=1}^{d}$ for the commuted order.  Consider also the
filter $\{\gamma_{1,s}(L)\}_{s=1}^{d}$, where
$\gamma_{2,s}(L)=\alpha_{1,s}(L)\beta_{1,s}(L)=1-(\alpha_{1,s}+\beta_{1,s})L+\alpha_{1,s}\beta_{1,s}L^2$
is the algebraic product of the polynomials for each
season. Table~\ref{Tab:AnExamplePeriodicFilter} shows the result of applying the filters
$\alpha_{1,s}(L)\beta_{1,s}(L)$, $\beta_{1,s}(L)\alpha_{1,s}(L)$ and $\gamma_{2,s}(L)$ to
$\{X_t\}$ . It can be seen from the table that the coefficients of $X_{t-2}$ for the three
cases are, in general, different. In particular, the sequential application of periodic
filters is non-commutative and the result is not obtained by simple multiplication of the
corresponding polynomials.

\begin{table}[H]
	\caption{An example for multiplication of periodic filters}
	\centering
	\label{Tab:AnExamplePeriodicFilter}
	\begin{tabular}{ll}
		\toprule               
		Filtering order &Result  \\
		\midrule
		first $\beta_{1,s}(L)$, then $\alpha_{1,s}(L)$ & 
		$X_t-(\alpha_{1,s}+\beta_{1,s})X_{t-1}+\alpha_{1,s}\beta_{1,s-1}X_{t-2}$\\
		first $\alpha_{1,s}(L)$, then $\beta_{1,s}(L)$ & 
		$X_t-(\alpha_{1,s}+\beta_{1,s})X_{t-1}+\beta_{1,s}\alpha_{1,s-1}X_{t-2}$\\
		$\gamma_{2,s}(L)$ & 
		$X_t-(\alpha_{1,s}+\beta_{1,s})X_{t-1}+\beta_{1,s}\alpha_{1,s}X_{t-2}$\\
		\bottomrule
	\end{tabular}
\end{table}

Note that the univariate representation in Eq (\ref{PARp Rep}) can be used to describe both
PAR and PIAR models. A useful method to determine whether the model is periodic
autoregressive or periodically integrated, is to research the set of polynomials
$\left\{\phi_{p,s}(z)\right\}_{s=1}^{d}$.  Note that, unlike the non-periodic case, this does
not amount to the study of the roots of the individual polynomials. Our study is based on
vector of seasons representation introduced in the following section.

\subsection{Vector of seasons representation}




An alternative way to study univariate periodic time series is to convert them into
multivariate ones by stacking the observations in each year in a vector.  To this end, let
$X_{[T,s]}$ be the observation for season $s$ of year $T$ and
$\X_T=(X_{[T,d]}, \dots, X_{[T,1]})^{'}$. A multivariate representation of $\{X_{t}\}$ is
$\{\X_{T}, \ T=1,2,\ldots\}$.  This idea was proposed originally by
\cite{gladyshev1961periodically} for the case of periodically stationary time series.
\cite{franses1994multivariate} used this representation extensively for the study of (mostly)
quarterly time series. He introduced the convenient term \emph{vector of quarters} (VQ)
representation of periodic time series for $d=4$.
\cite{boshnakov2009generation} proposed the term \emph{vector of seasons} (VS) for general
values of~$d$. Note that for each fixed $s \in [1,d]$ the subseries
$\left\{X_{[T,s]}, T=1,2,\dots\right\}$ is the seasonal component (corresponding to season
$s$) of the univariate periodic time series.


The VS representation of the model given by Eq~\eqref{PARp Rep} is
\begin{equation}
\label{PARp VSRep}
\begin{aligned}
&\Phi_0\X_T=\sum_{i=1}^{P}\Phi_i\X_{T-i}+\eps_T,
&T=1,2,\dots,
\end{aligned}
\end{equation}
where $\eps_T=(\eps_{[T,d]}, \dots, \eps_{[T,1]})^{'}$ is the vector of seasons form of $\eps_t$; $P=1+[(p-1)/d]$ with $[\cdot]$ is the integer function; $\Phi_0$ and $\Phi_i$ are

\begin{equation*}
\begin{aligned}
(\Phi_0)_{jk}
&=
\begin{cases}
1  & j=k \\
0 & j>k\\
-\phi_{j-i,d-i+1} & j<k\\
\end{cases},\\
(\Phi_i)_{jk}&=\phi_{k+di-j,d-j+1},
\quad i=1,\dots,P,
\end{aligned}
\end{equation*}
for $j,k=1,\dots,d$. Notice that notation $(M)_{jk}$ means the $(j,k)$-th element of matrix $M$. 

In order to distinguish between a PAR or a PIAR model, we consider the characteristic equation
of Eq (\ref{PARp VSRep}), see below:

\begin{equation*}
|\Phi(z)|=|\Phi_0-\Phi_1z-\cdots-\Phi_pz^{p}|=0.
\end{equation*}

When the roots of the characteristic equation are outside the unit circle, the VS process
$\left\{\X_T\right\}$ in Eq (\ref{PARp VSRep}) is stationary, the corresponding univariate
process $\left\{X_t\right\}$ is periodically stationary, and Eq (\ref{PARp Rep}) is a
periodic autoregression of order $p$, namely PAR$(p)$. In contrast, when there is at least
one root on the unit circle (which is called as unit root), $\left\{\X_T\right\}$ in Eq
(\ref{PARp VSRep}) is integrated, the corresponding univariate process $\left\{X_t\right\}$
is periodically integrated, and Eq (\ref{PARp Rep}) is a periodically integrated
autoregression of order $p$, namely PIAR$(p)$.

Previous study see \cite{franses1996multistep}, \cite{franses2004periodic} and \cite{franses2005forecasting} provided both theoretical and empirical analysis of forecasts for quarterly periodic models by using the VQ representation. Their methods can be extended to general cases and here, we derive explicit expressions for $H$-year ahead forecasts (with $H \geq 1$) and forecast error variances for both PAR$(p)$ and PIAR$(p)$ models with $p \leq d$. 

Let $\left\{X_t, t=1,2, \dots,n\right\}$ be a periodic time series with period $d$ which is represented in Eq (\ref{PARp Rep}), and let $\left\{\X_T, T=1,2,\dots,N\right\}$ be the corresponding VS process where $N=n/d$ is the final year within the observations. The $H$-year ahead forecasts are generated from year $N$ onwards, denoted by $\hat{\X}_{N+H}$. Based on the VS representation in Eq (\ref{PARp VSRep}), we derive $H$-year ahead forecast, forecast error and forecast error variance:

\begin{equation*}
\begin{aligned}
&\hat{\X}_{N+H}=(\Phi_0^{-1}\Phi_1)^{H}\X_N,\\
&\X_{N+H}-\hat{\X}_{N+H}=\sum_{h=0}^{H-1}\left[(\Phi_0^{-1}\Phi_1)^{h}\Phi_0^{-1}\right]\eps_{N+H-h},\\
&\E[(\X_{N+H}-\hat{\X}_{N+H})(\X_{N+H}-\hat{\X}_{N+H})^{'}]=\sum_{h=0}^{H-1}(\Phi_0^{-1}\Phi_1)^{h}\Phi_0^{-1}\Sigma_{\eps}\left[(\Phi_0^{-1}\Phi_1)^{h}\Phi_0^{-1}\right]^{'},
\end{aligned}
\end{equation*}
where $\Sigma_{\eps}=\text{diag}(\sigma_d^2,\dots,\sigma_1^2)$, and $(\Phi_0^{-1}\Phi_1)^{0}$ is defined to be an identity matrix. \cite{franses1996multistep} proved that for a quarterly PIAR(1) model, the matrix $\Phi_0^{-1}\Phi_1$ is idempotent, i.e. $(\Phi_0^{-1}\Phi_1)^{h}=\Phi_0^{-1}\Phi_1$ for $h=1,2,\dots$. In fact, this property holds for PIAR(1) models with any period. Therefore, the $H$-year ahead forecast of PIAR(1) models remains the same, i.e. $\hat{\X}_{N+H}=(\Phi_0^{-1}\Phi_1)\X_N$ for $H=1,2,\dots$, and the corresponding forecast error variance reduces to $(H-1)(\Phi_0^{-1}\Phi_1)\Phi_0^{-1}\Sigma_{\eps}\left[(\Phi_0^{-1}\Phi_1)\Phi_0^{-1}\right]^{'}+\Phi_0^{-1}\Sigma_{\eps}{(\Phi_0^{-1})}^{'}$.

\subsection{Multi-companion representation}

Lastly, we introduce a multi-companion representation of the model Eq (\ref{PARp Rep}). This representation is developed by using the multi-companion matrix \citep[see][]{boshnakov2002multi}, and is significantly useful for our later analysis on periodic integration.

\cite{boshnakov2009generation} proposed a Markov form of univariate periodic time series models in Eq (\ref{PARp Rep}), which is:

\begin{equation}
\label{PARp ComRep}
\begin{aligned}
&\X_t=A_t\X_{t-1}+\bm{E}_t,
&t=1,2,\dots,
\end{aligned}
\end{equation}
where $\X_t=(X_t,X_{t-1},\dots,X_{t-m+1})^{'}$ and $\bm{E}_t=(\eps_t,0,\dots,0)^{'}$ are $m$-dimensional vector with $m=\max(p,d)$, and $A_t$ is an $m \times m$ companion matrix such that

\begin{equation*}
A_t=
\begin{pmatrix}
\phi_{1,t}&\phi_{2,t}&\dots&\phi_{m-1,t}&\phi_{m,t}\\
1&0&\dots&0&0\\
0&1&\dots&0&0\\
\vdots&\vdots&\ddots&\vdots&\vdots\\
0&0&\dots&1&0
\end{pmatrix},
\end{equation*}
with $\phi_{i,t}=0$ for $i>p$. Eq (\ref{PARp ComRep}) is also called as the companion representation of Eq (\ref{PARp Rep}). Note that the companion matrix $A_t$ in Eq (\ref{PARp ComRep}) is $d$-periodic in time $t$ such that $A_t=A_{t+d}$, hereby, it is sufficient to consider $A_1,\dots,A_d$ only. 

Given the $[T,s]$ notation mentioned in Section \ref{Introduction}, we replace time $t$ with $[T,s]$ and the companion representation Eq (\ref{PARp ComRep}) changes to

\begin{equation}
\label{PARp ComRepv1}
\begin{aligned}
&\X_{[T,s]}=A_s\X_{[T,s-1]}+\bm{E}_{[T,s]},
&[T,s]=1,2,\dots.
\end{aligned}
\end{equation}

Finally, by iterating Eq (\ref{PARp ComRepv1}), we have the multi-companion representation \citep[see][]{boshnakov2009generation}

\begin{equation}
\label{PARp MCRep}
\begin{aligned}
&\X_T=\F_d\X_{T-1}+\bm{u}_T,
&T=1,2,\dots,
\end{aligned}
\end{equation}
where $\X_T=(X_{[T,d]},\dots,X_{[T,d]-m+1})^{'}$, $\F_d=A_d A_{d-1}\cdots A_1$, and $\bm{u}_T=\bm{E}_{[T,d]}+\sum_{i=1}^{d-1}\prod_{j=1}^{i}A_{d-j+1}\bm{E}_{[T,d]-i}$. Note that $\F_d$ in Eq (\ref{PARp MCRep}) is a product of $d$ companion matrices, and is a multi-companion matrix with companion order $d$ \citep[see][Corollary. 3.2]{boshnakov2002multi}. In addition, $\F_d$ is a constant matrix independent of time $t$. Importantly, the whole process by expressing Eq (\ref{PARp Rep}) into (\ref{PARp MCRep}) demonstrates that the multi-companion matrix $\F_d$ in multi-companion representation Eq (\ref{PARp MCRep}) determines completely the properties of the corresponding periodic filter in univariate representation Eq (\ref{PARp Rep}). 

Furthermore, we find that the disturbance term $\bm{u}_T$ in Eq (\ref{PARp MCRep}) can be expressed as a linear combination of the periodic white noise term $\eps_t$ in Eq (\ref{PARp Rep}), such that $\bm{u}_T=\Omega\eps_T$ where $\eps_T=(\eps_{[T,d]}, \eps_{[T,d]-1}, \dots, \eps_{[T,d]-m+1})^{'}$ and $\Omega$ is a matrix defined by

\begin{equation}
\label{Omega}
\Omega=[e_1, (A_d)_{\bullet 1}, (A_dA_{d-1})_{\bullet 1}, \dots, (A_dA_{d-1}\cdots A_2)_{\bullet 1}, 0_{m \times (m-d)}],
\end{equation}
where $e_1$ is a unit vector with its first component equal to 1 and all other components equal to 0, and the notation $()_{\bullet 1}$ stands for the first column of a matrix. Particularly, when $m=d$, $\Omega$ is an upper triangular matrix with main diagonal elements equal to one.

It is interesting to investigate the characteristic equation of the multi-companion representation in Eq (\ref{PARp MCRep}), which is $|\mathit{I}-\F_dz|=0$ where $\mathit{I}$ is an $m \times m$ identity matrix. In fact, the roots of this characteristic equation are the reciprocal of the eigenvalues of the multi-companion matrix $\F_d$. When the roots of the characteristic equation are all outside the unit circle, which is equivalent to when all the eigenvalues of $\F_d$ have moduli strictly less than one, then Eq (\ref{PARp MCRep}) is a periodic autoregression of order $p$. In contrast, when there is at least one root of the characteristic equation on the unit circle, which is equivalent to there is at least one unit eigenvalue of $\F_d$, then Eq (\ref{PARp MCRep}) is a periodically integrated autoregression. This indicates that the study of the eigen information of the multi-companion matrix can be helpful to examine the properties of periodic models. 

To continue with $H$-year ahead forecast, we still set $\left\{X_t,t=1,2,\dots,n\right\}$ as the observations and $N$ is the final year within the observations such that $N=n/d$. Based on the multi-companion representation in Eq (\ref{PARp MCRep}), we derive $H$-year ahead forecast, forecast error and forecast error variance as:

\begin{equation*}
\begin{aligned}
&\hat{\X}_{N+H}=\F_d^{H}\X_{N},\\
&\X_{N+H}-\hat{\X}_{N+H}=\sum_{h=0}^{H-1}\F_d^{h}\bm{u}_{N+H-h},\\
&\E[(\X_{N+H}-\hat{\X}_{N+H})(\X_{N+H}-\hat{\X}_{N+H})^{'}]=\sum_{h=0}^{H-1}\F_d^{h}\Sigma_u(\F_d^{h})^{'},
\end{aligned}
\end{equation*}
where $\Sigma_u=\Omega\Sigma_{\eps}\Omega^{'}$ and $\F_d^{0}$ is defined to be an identity matrix. We will show in next section that $\F_d$ can be expressed in its Jordan canonical form, i.e. $\F_d=XJX^{-1}$, and hereby, we have $\F_d^{h}=XJ^{h}X^{-1}$ for $h=1,2,\dots$. It is noticeable that if $\F_d$ is diagonalizable and all the eigenvalues of $\F_d$ are either 1 or 0, then $\F_d$ is idempotent which results in the $H$-year ahead forecast of the model remains the same, i.e. $\hat{\X}_{N+H}=\F_d\X_N$ for $H=1,2,\dots$, and the corresponding forecast error variance reduces to $\Sigma_u+(H-1)\F_d\Sigma_u(\F_d)^{'}$. This situation can happen when the corresponding model is a periodically integrated autoregression and we will elaborate on this further in later sections.



\section{Multi-companion method}
\label{Multi-companion method}

In this section, we introduce a multi-companion method which is used to investigate the periodic models, particular for periodically integrated autoregressive models. The multi-companion method is based on the multi-companion representation Eq (\ref{PARp MCRep}) and the eigen information of the multi-companion matrix.

It is useful to review some important properties of multi-companion matrices before introducing the multi-companion method. For a given $d$-companion matrix $\F_d$ with dimension $m$, it can be decomposed into its Jordan canonical form $\F_d=XJX^{-1}$, where $X$ is the similarity matrix consisting of the eigenvectors of $\F_d$, and $J$ is the Jordan matrix whose diagonal elements are the eigenvalues of $\F_d$. In addition, let $\lambda_i$ be the eigenvalue and $x_i$ be the corresponding eigenvector of $\F_d$ for $i=1,\dots,m$. The first important property is that each eigenvector $x_i$ of $\F_d$ is determined uniquely by its first (or any consecutive) $d$ elements and the corresponding eigenvalue $\lambda_i$, see \cite{boshnakov2009generation}. Hereby, we define the first $d$ elements of the eigenvector $x_i$ to be the \emph{seed-parameters}, which are denoted by $\co{i}{j}$ for $j=1,\dots,d$. The vector $c_i=(\co{i}{1},\dots,\co{i}{d})^{'}$ which consists of the $d$ seed-parameters is defined to be a seed-vector. The second property is that the eigenvectors corresponding to zero eigenvalues of $\F_d$ are some appropriate standard basis vectors, see \citet[][Lemma.1]{boshnakov2009generation}. This property is a particular example of specifying eigenvectors corresponding to zero eigenvalues and will be employed in Section \ref{Multi-companion method applied to PIAR model}.

Considering the aforementioned properties, it seems appropriate to explore the potential use of the eigen information of the multi-companion matrix for periodic time series models. Back to the multi-companion representation Eq (\ref{PARp MCRep}) of a periodic model, substituting $\F_d$ by its Jordan canonical form and then left-multiplying $X^{-1}$ to its both sides gives

\begin{equation*}
\label{PARp MCRepv1}
X^{-1}\X_T=JX^{-1}\X_{T-1}+X^{-1}\bm{u}_T.
\end{equation*}

By defining two processes $\Z_T=X^{-1}\X_T$ and $\W_T=X^{-1}\bm{u}_T$, we can rewrite the above equation into:

\begin{align}
\label{XT define}
\X_T&=X\Z_T,\\
\label{ZT define}
\Z_T&=J\Z_{T-1}+\W_T,
\end{align}
where $\Z_T=(Z_T^{(1)}, \dots, Z_T^{(m)})^{'}$ is an $m$-dimensional  process and $\W_T=(W_T^{(1)}, \dots, W_T^{(m)})^{'}$ is an $m$-dimensional white noise. It is then straightforward to see that Eq (\ref{XT define}) uses the similarity matrix of $\F_d$ as a coefficient matrix linking process $\Z_T$ with $\X_T$, and Eq (\ref{ZT define}) is in the vector autoregression form where the Jordan matrix of $\F_d$ is regarded as an autoregressive coefficient matrix. Therefore, the eigen information of $\F_d$ plays an important role in analysing periodic time series models. The subsequent two subsections will provide an in-depth analysis of the roles of similarity and Jordan matrices of $\F_d$ respectively.

\subsection{The role of similarity matrix}

Considering the role of similarity matrix, we concentrate on Eq (\ref{XT define}) which shows that $\X_T$ is represented in terms of the similarity matrix and the vector process $\Z_T$. Expanding Eq (\ref{XT define}) and only considering the first $d$ elements gives

\begin{equation*}
\begin{pmatrix}
X_{[T,d]}\\
X_{[T,d-1]}\\
\vdots\\
X_{[T,1]}
\end{pmatrix}
=
\begin{pmatrix}
c_1^{(1)}&c_2^{(1)}&\dots&c_m^{(1)}\\
c_1^{(2)}&c_2^{(2)}&\dots&c_m^{(2)}\\
\vdots&\vdots&\dots&\vdots\\
c_1^{(d)}&c_2^{(d)}&\dots&c_m^{(d)}\\
\end{pmatrix}
\begin{pmatrix}
Z_T^{(1)}\\
Z_T^{(2)}\\
\vdots\\
Z_T^{(d)}\\
\vdots\\
Z_T^{(m)}
\end{pmatrix},
\end{equation*}
which implies

\begin{equation*}
\begin{aligned}
X_{[T,d]}&=c_1^{(1)}Z_T^{(1)}+c_2^{(1)}Z_T^{(2)}+\cdots+c_m^{(1)}Z_T^{(m)},\\
X_{[T,d-1]}&=c_1^{(2)}Z_T^{(1)}+c_2^{(2)}Z_T^{(2)}+\cdots+c_m^{(2)}Z_T^{(m)},\\
&\; \vdots\\
X_{[T,1]}&=c_1^{(d)}Z_T^{(1)}+c_2^{(d)}Z_T^{(2)}+\cdots+c_m^{(d)}Z_T^{(m)}.\\
\end{aligned}
\end{equation*}

The above systems can be further summarized as:

\begin{equation}
\begin{aligned}
\label{Role of similarity}
&X_{[T,s]}=\sum_{i=1}^{m}c_i^{(d-s+1)}Z_T^{(i)},
&T=1,2,\dots,
\end{aligned}
\end{equation}
for season $s \in [1,d]$. Eq (\ref{Role of similarity}) shows that the seasonal component $\left\{X_{[T,s]}, T=1,2,\dots\right\}$ at season $s$ can be viewed as a linear combination of elements of $\left\{\Z_T, T=1,2,\dots\right\}$, and the coefficient $c_i^{(d-s+1)}$ (the seed-parameter) is interpreted as the ``strength" of the influence of $Z_T^{(i)}$ on the $s$-th season. It indicates that if one of $Z_T^{(i)}$ processes is a random walk and its corresponding coefficients for each season are non-zero, then this random walk will be a common stochastic trend driving the entire process. In addition, when some $Z_T^{(i)}$ processes need to be eliminated from one seasonal component at season $s$, we can directly set their corresponding coefficients to be zero. We will show later how Eq (\ref{Role of similarity}) contributes to analysing PIAR models with a single, two and multiple unit roots.

\subsection{The role of Jordan matrix}

We concentrate on Eq (\ref{ZT define}) when taking into account the significance of the Jordan matrix. It is evident that if all the diagonal elements of the Jordan matrix $J$ have modulus strictly than one, the process $\Z_T$ is stationary and Eq (\ref{ZT define}) is a vector autoregression of order one (VAR(1)). In our paper, we pay more attention to the cases when Jordan matrix $J$ has at least one diagonal element equal to one.

Let $\eigunit$ be the unit eigenvalue of $\F_d$, such that $\eigunit=1$. We use notations $\Am(\eigunit)$ and $\Gm(\eigunit)$ to describe the algebraic and geometric multiplicities of $\eigunit$ respectively. In addition, let $\Junit$ be the unit Jordan matrix which consists of unit eigenvalues of $\F_d$, such that

\begin{equation}
\label{Junit}
\Junit= \text{diag}({J^{(1)}_{\text{unit}}, J^{(2)}_{\text{unit}}, \cdots, J^{(g)}_{\text{unit}}}),
\end{equation}
where $J^{(k)}_{\text{unit}}$ is the $k$-th unit Jordan block of dimension $r_k$ for $k=1,\dots,g$; $g$ is the number of unit Jordan blocks which is determined by the geometric multiplicity of $\eigunit$, namely $g=\text{Gm}(\eigunit)$; and the sum of the dimension of each unit Jordan block equals the algebraic multiplicity, namely $\sum_{k=1}^{g}r_k=\text{Am}(\eigunit)$. Moreover, the general form of each unit Jordan block is

\begin{equation*}
J^{(k)}_{\text{unit}}=
\begin{pmatrix}
1&1&&\\
&1&\ddots&\\
&&\ddots&1\\
&&&1\\
\end{pmatrix}
\in 
\R^{r_k \times r_k},
\quad
k=1,\dots,g.
\end{equation*}

Throughout our paper, we suppose that the diagonal elements of Jordan matrix is arranged in descending order such that if the unit Jordan matrix exists, then it is arranged in the top-left corner of $J$.

In particular, we assume there are $m_1$ unit eigenvalues of $\F_d$ where $m_1 \in [1,m]$, and the remaining $(m-m_1)$ eigenvalues of $\F_d$ have modulus strictly less than one. Under this assumption, the Jordan matrix $J$ of $\F_d$ can be expressed as $J=\text{diag}(\Junit,\Lambda_{m-m_1})$ where $\Junit$ is defined by Eq (\ref{Junit}) with the sum of the dimension of $g$ unit Jordan blocks equal to $m_1$, namely $\sum_{k=1}^{g}r_k=m_1$, and $\Lambda_{m-m_1}$ corresponds to the stationary part whose diagonal elements have moduli strictly less than one. As indicated from Eq (\ref{ZT define}), each unit Jordan block $J^{(k)}_{\text{unit}}$ corresponds to $r_k$ elements of $\Z_T$ process, and the highest integration order of the corresponding elements of $\Z_T$ process is determined by the dimension of the unit Jordan block, namely $r_k$. The first unit Jordan block $J^{(1)}_{\text{unit}}$, for example, corresponds to the first $r_1$ elements of $\bm{Z}_T$, where $Z_T^{(1)} \sim \text{I}(r_1)$, $Z_T^{(2)} \sim \text{I}(r_1-1)$, $\dots$, $Z_T^{(r_1)} \sim \text{I}(1)$. Obviously, $Z_T^{(1)}$ has the highest integration order among the first $r_1$ elements of $\Z_T$, which is exactly equal to the dimension of $J^{(1)}_{\text{unit}}$. Similarly, the $k$-th unit Jordan block $J^{(k)}_{\text{unit}}$ for each $k=2,\dots,g$ corresponds to 

\begin{equation*}
\begin{aligned}
\label{ZT integration order}
Z_T^{(\sum_{i=1}^{k-1}r_i+1)} & \sim \text{I}(r_k),\\
Z_T^{(\sum_{i=1}^{k-1}r_i+2)} & \sim \text{I}(r_k-1),\\
&\; \vdots\\
Z_T^{(\sum_{i=1}^{k-1}r_i+r_k)} & \sim \text{I}(1),
\end{aligned}
\end{equation*}
where $Z_T^{(\sum_{i=1}^{k-1}r_i+1)}$ has the highest integration order which is exactly equal to the dimension of the $k$-th unit Jordan block, $r_k$.

To illustrate how the $Z_T^{(i)}$ processes drive the series $\left\{X_t\right\}$ periodically non-stationary, we give two examples below.

The first example applies when $\Am(\eigunit)=\Gm(\eigunit)=m_1$. Under this condition, the unit Jordan matrix is exactly an $m_1 \times m_1$ identity matrix and Eq (\ref{ZT define}) is expanded as

\begin{equation*}
\label{ZT Am=Gm}
\begin{pmatrix}
Z_T^{(1)}\\
Z_T^{(2)}\\
\vdots\\
Z_T^{(m_1)}\\
\tilde{Z}_T
\end{pmatrix}
=
\begin{pmatrix}
1&&&&\\
&1&&&\\
&&\ddots&&\\
&&&1&\\
&&&&\Lambda_{m-m_1}
\end{pmatrix}
\begin{pmatrix}
Z_{T-1}^{(1)}\\
Z_{T-1}^{(2)}\\
\vdots\\
Z_{T-1}^{(m_1)}\\
\tilde{Z}_{T-1}
\end{pmatrix}
+
\begin{pmatrix}
W_T^{(1)}\\
W_T^{(2)}\\
\vdots\\
W_T^{(m_1)}\\
\tilde{W}_T\\
\end{pmatrix},
\end{equation*}
where $\tilde{Z}_T=(Z_{T}^{(m_1+1)},\dots,Z_T^{(m)})^{'}$ and $\tilde{W}_T=(W_{T}^{(m_1+1)},\dots,W_T^{(m)})^{'}$ are the two vector processes of dimension $m-m_1$. The above matrix form implies

\begin{equation*}
\label{tilde ZT Am=Gm}
\begin{aligned}
&Z_T^{(i)}=Z_{T-1}^{(i)}+W_T^{(i)},
\quad i=1,\dots,m_1;\\
&\tilde{Z}_T=\Lambda_{m-m_1}\tilde{Z}_{T-1}+\tilde{W}_T.
\end{aligned}
\end{equation*}
where the first equation is a random walk such that $Z_T^{(i)} \sim \text{I}(1)$ for each $i \in [1,m_1]$, and the second equation is a vector autoregression such that $Z_T^{(i)} \sim \text{I}(0)$ for each $i \in [m_1,m]$. As indicated from Eq (\ref{Role of similarity}), these $Z_T^{(i)}$ processes have impact on the seasonal component of $\left\{\X_T\right\}$, such that

\begin{equation*}
\begin{aligned}
X_{[T,s]}=\sum_{i=1}^{m_1}c_i^{(d-s+1)}Z_T^{(i)}+\sum_{i=m_1+1}^{m}c_i^{(d-s+1)}Z_T^{(i)},
\quad
T=1,2,\dots,
\end{aligned}
\end{equation*}
for each $s \in [1,d]$. Therefore, it is concluded that $Z_T^{(1)},\dots,Z_T^{(m_1)}$ are the common stochastic trends (random walks) driving each seasonal component $\left\{X_{[T,s]}, T=1,2,\dots\right\}$ non-stationary. Moreover, for each season $s$, if there is at least one seed-parameter $c_i^{(d-s+1)}$ non-zero for any $i \in [1,m_1]$, then the seasonal component will be integrated of order one, denoted by $\left\{X_{[T,s]}, T=1,2,\dots\right\} \sim \text{I}(1)$. We will explain later that this example corresponds to the case where the series $\left\{X_t\right\}$ is periodically integrated of order one.

The second example is considered when $\Am(\eigunit)=m_1>\Gm(\eigunit)=1$. Under this condition, the unit Jordan matrix is exactly a unit Jordan block of dimension $m_1 \times m_1$, and Eq (\ref{ZT define}) is expanded as

\begin{equation*}
\begin{pmatrix}
Z_T^{(1)}\\
Z_T^{(2)}\\
\vdots\\
Z_T^{(m_1)}\\
\tilde{Z}_T
\end{pmatrix}
=
\begin{pmatrix}
1&1&&&\\
&1&\ddots&&\\
&&\ddots&1&\\
&&&1&\\
&&&&\Lambda_{m-m_1}
\end{pmatrix}
\begin{pmatrix}
Z_{T-1}^{(1)}\\
Z_{T-1}^{(2)}\\
\vdots\\
Z_{T-1}^{(m_1)}\\
\tilde{Z}_{T-1}
\end{pmatrix}
+
\begin{pmatrix}
W_T^{(1)}\\
W_T^{(2)}\\
\vdots\\
W_T^{(m_1)}\\
\tilde{W}_T\\
\end{pmatrix},
\end{equation*}
where the top-left corner is the unit Jordan block. The above equation implies 

\begin{equation*}
\begin{aligned}
\label{ZT Am>Gm}
&Z_T^{(i)}=Z_{T-1}^{(i)}+Z_{T-1}^{(i+1)}+W_T^{(i)},&i=1,\dots,m_1-1;\\
&Z_T^{(m_1)}=Z_{T-1}^{(m_1)}+W_T^{(m_1)};\\
&\tilde{Z}_T=\Lambda_{m-m_1}\tilde{Z}_{T-1}+\tilde{W}_T.
\end{aligned}
\end{equation*}
where the first two equations lead to $Z_T^{(i)} \sim \text{I}(m_1-i+1)$ for $i=1,\dots,m_1$, and the last equation indicates $Z_T^{(i)} \sim \text{I}(0)$ for $i=m_1+1,\dots,m$. In this example, $Z_T^{(1)}$ has the largest integration order such that $Z_T^{(1)} \sim \text{I}(m_1)$, which ensures the seasonal components are integrated of order $m_1$ if $\co{1}{d-s+1}$ is non-zero for any $s$, denoted by $\left\{X_{[T,s]}, T=1,2,\dots\right\} \sim \text{I}(m_1)$. We will explain later that the second example corresponds to the case where the series $\left\{X_t\right\}$ is periodically integrated of order $m_1$. Note that this case with periodic integration order lager than one has not been discussed by \cite{boswijk1996unit} or other current literatures. 

In conclusion, we have noticed that the largest dimension of unit Jordan blocks will affect the integration order of seasonal components $\left\{X_{[T,s]}, T=1,2,\dots\right\}$, and in turn, will influence the periodic integration order of the entire process $\left\{X_t, t=1,2,\dots\right\}$. Therefore, it is reasonable to consider using the property of Jordan matrix to have the following definition for periodic integration.

\begin{definition}(Periodic integration).
	\label{Def:Periodic integration}
	Let $\left\{X_t\right\}$ be a series defined by Eq (\ref{PARp Rep}) with multi-companion representation Eq (\ref{PARp MCRep}). Suppose $\F_d$ in Eq (\ref{PARp MCRep}) has at least one unit eigenvalue and its corresponding unit Jordan matrix $\Junit$ is represented in Eq (\ref{Junit}). Then, $\left\{X_t\right\}$ is said to be periodically integrated of order $r$, denoted by $X_t \sim \text{PI}(r)$, if the largest dimension of the unit Jordan blocks is $r=\max(r_1,\dots,r_g)$, where $r_k$ for $k=1,\dots,g$ is the dimension of $k$-th unit Jordan block.
\end{definition}

The previous study \citet[][Def. 1]{boswijk1996unit} has introduced a definition for (quarterly) periodic integration of order one. They stated that if the VQ representation of the model (see Eq (\ref{PARp VSRep}) by setting $d=4$) has a single unit root and if all the seasonal components of $\left\{X_t\right\}$ are integrated of order one, namely $\left\{X_{[T,s]}, T=1,2,\dots\right\} \sim \text{I}(1)$ for any $s$, then $\left\{X_t\right\}$ is said to be periodically integrated of order one, denoted by $X_t \sim \text{PI}(1)$. It is easy to show that \citet[][Def. 1]{boswijk1996unit} is a special case of our Definition \ref{Def:Periodic integration}. Recall the first example aforementioned, let $m_1=1$ which ensures there is a single unit root of $\left\{X_t\right\}$, and moreover, we have shown that in this case each seasonal component of $\left\{X_t\right\}$ is integrated of order one. Therefore, two conditions in \citet[][Def. 1]{boswijk1996unit} are satisfied and we have the conclusion that $X_t \sim \text{PI}(1)$. On the other hand, it is obvious to have $X_t \sim \text{PI}(1)$ according to our Definition \ref{Def:Periodic integration} when setting $m_1=1$ in the first example above, as in this case the largest dimension of the unit Jordan block is one. 

In general, Definition \ref{Def:Periodic integration} developed by using our multi-companion method extends the previous study of \citet[][Def. 1]{boswijk1996unit}. Furthermore, it is obvious to deduce that the two aforementioned examples correspond to the cases where $X_t \sim \text{PI}(1)$ and $X_t \sim \text{PI}(m_1)$ respectively, according to Definition \ref{Def:Periodic integration}. For future reference, we use notation $\text{PI}(0)$ to describe the periodically stationary process.

\section{Multi-companion method applied to PIAR models}
\label{Multi-companion method applied to PIAR model}

Section \ref{Multi-companion method} introduces the multi-companion method which relies on the eigen information of the multi-companion matrix. In this section, we will demonstrate how the multi-companion method is applied to analyse PIAR models with a single, two and multiple unit roots. For each case, we find a periodically integrated filter which transforms the periodically integrated series into periodically stationary. In addition, we derive the representation of the the parameters of the periodically integrated filter in terms of the eigen information of the multi-companion matrix. Based on the parametrization process, we propose an innovative estimation method to estimate the parameters of the PIAR models.

\subsection{A single unit root}

Let $\left\{X_t\right\}$ be the series generated by Eq (\ref{PARp Rep}) with multi-companion representation Eq (\ref{PARp MCRep}). Suppose $\F_d$ in Eq (\ref{PARp MCRep}) has a single unit eigenvalue. Under this assumption, the corresponding series $\left\{X_t\right\}$ is periodically integrated of order one, denoted by $X_t \sim \text{PI}(1)$, according to Definition \ref{Def:Periodic integration}. Subsequently, Eq (\ref{PARp Rep}) is a $\PIAR{1}{p}$ model with a singe unit root. 

For simplicity, we first consider $\F_d$ in Eq (\ref{PARp MCRep}) has a single unit eigenvalue and all the other eigenvalues are zero. In this case, the Jordan canonical form of $\F_d$ is 

\begin{equation}
\label{Fd SingleUnit Rep}
\begin{aligned}
\F_d&=XJX^{-1}\\
&=
\begin{pmatrix}
\co{1}{1}&0&\dots&0&0\\
\co{1}{2}&1&\dots&0&0\\
\vdots&\vdots&\ddots&\vdots&\vdots\\
\co{1}{d-1}&0&\dots&1&0\\
\co{1}{d}&0&\dots&0&1\\
\end{pmatrix}
\begin{pmatrix}
1&&&&\\
&0&&&\\
&&\ddots&&\\
&&&0&\\
&&&&0\\
\end{pmatrix}
\begin{pmatrix}
\co{1}{1}&0&\dots&0&0\\
\co{1}{2}&1&\dots&0&0\\
\vdots&\vdots&\ddots&\vdots&\vdots\\
\co{1}{d-1}&0&\dots&1&0\\
\co{1}{d}&0&\dots&0&1\\
\end{pmatrix}^{-1}\\
&=
\begin{pmatrix}
1&0&\dots&0&0\\
\frac{\co{1}{2}}{\co{1}{1}}&0&\dots&0&0\\
\vdots&\vdots&\ddots&\vdots&\vdots\\
\frac{\co{1}{d-1}}{\co{1}{1}}&0&\dots&0&0\\
\frac{\co{1}{d}}{\co{1}{1}}&0&\dots&0&0\\
\end{pmatrix},
\end{aligned}
\end{equation}
where the eigenvectors corresponding to zero eigenvalues are some appropriate standard basis vectors \citep[see][Lemma. 1]{boshnakov2009generation}. Considering the role of Jordan matrix, see Eq (\ref{ZT define}), it implies that $Z_T^{(1)}$ is a random walk which is the only non-stationary part among all the elements of $\Z_T$ process, and $Z_T^{(i)}, i=2,\dots,d$ are white noise. On the other hand, the role of similarity matrix, see Eq (\ref{XT define}), shows each seasonal component of $\left\{\X_T\right\}$ at year $T=1,2,\dots$ can be expressed as

\begin{equation*}
\begin{aligned}
X_{[T,d]}&=\co{1}{1}Z_T^{(1)},\\
X_{[T,d-1]}&=\co{1}{2}Z_T^{(1)}+Z_T^{(2)},\\
&\vdots\\
X_{[T,1]}&=\co{1}{d}Z_T^{(1)}+Z_T^{(d)},\\
\end{aligned}
\end{equation*}
where $Z_T^{(1)}$ is the common stochastic trend driving each seasonal component integrated of order one. In such a situation, a periodic filter $(1-\alpha_sL)$ is introduced to remove the non-stationary part $Z_T^{(1)}$ from each seasonal component $\left\{X_{[T,s]}, T=1,2,\dots\right\}$, where $\alpha_s$ are determined by

\begin{equation}
\begin{aligned}
\label{alphas SingleUnit}
\alpha_s=\frac{\co{1}{d-s+1}}{\co{1}{d-s+2}},
\quad 
s=1,\dots,d,
\end{aligned}
\end{equation}
with $\co{1}{d+1}=\co{1}{1}$. 

It is noticeable that $\alpha_s$ in Eq (\ref{alphas SingleUnit}) automatically satisfies the restriction $\prod_{s=1}^{d}\alpha_s=1$. The previous study \cite{osborn1988seasonality} took this restriction as the defining property of a unit root periodic filter by restricting $d=4$ for quarterly cases. Obviously, our multi-companion method extends the quarterly cases to general situations. For future reference, we use the term periodically integrated filter (\emph{PI-filter}) to describe the periodic filters which are used to remove the unit roots in the process. Particularly, when the PI-filter is with order one, namely $(1-\alpha_sL)$ where $\alpha_s$ satisfies the restriction $\prod_{s=1}^{d}\alpha_s=1$, we call it as a unit PI-filter.

In general, when $\F_d$ in Eq (\ref{PARp MCRep}) has a single unit eigenvalue and all other eigenvalues have moduli strictly less than one, model in Eq (\ref{PARp Rep}) can be rewritten as

\begin{equation*}
\begin{aligned}
&\psi_{p-1,s}(L)(1-\alpha_sL)X_t=\eps_t,
&t=1,2,\dots,
\end{aligned}
\end{equation*}
where $\psi_{p-1,s}(L)$ is a periodic autoregressive filter with order $p-1$, and $(1-\alpha_sL)$ is a unit PI-filter where $\alpha_s$ are determined by Eq (\ref{alphas SingleUnit}), automatically satisfying the non-linear restriction $\prod_{s=1}^{d}\alpha_s=1$.

\subsection{Two unit roots}

Before investigating the two unit roots cases, we first illustrate some key terminology. We
use the term \emph{simple unit eigenvalues} to denote unit eigenvalues of the multi-companion matrix that are in different unit Jordan blocks. In other words,  the algebraic multiplicity of the unit eigenvalue is equal to its geometric multiplicity, namely $\Am(\eigunit)=\Gm(\eigunit)$. Consequently, the resulting unit roots in the model Eq (\ref{PARp MCRep}) are referred to as \emph{simple unit roots}.

Conversely, we use the term \emph{chained unit eigenvalues} to describe unit eigenvalues of the multi-companion matrix that are contained within the same unit Jordan block. In other words, the chained unit eigenvalues are in a same Jordan chain and $\Am(\eigunit) > \Gm(\eigunit)$ holds. Correspondingly, the unit roots generated in the model Eq (\ref{PARp MCRep}) are termed \emph{chained unit roots}.

In this subsection, we assume that $\F_d$ in Eq (\ref{PARp MCRep}) has two unit eigenvalues. Under the assumption, the series generated by this $\F_d$ can have either two simple or two chained unit roots. Consequently, the model in Eq (\ref{PARp MCRep}) can either be $\PIAR{1}{p}$ or $\PIAR{2}{p}$, depending on if the two unit eigenvalues are in a same Jordan block. We will show that the PI-filters employed to remove two simple unit roots differ from those used to eliminate the two chained unit roots. Specifically, we will present parametrization results for the PI-filters in both cases.

\subsubsection{Two simple unit roots}

The first case occurs when $\F_d$ in Eq (\ref{PARp MCRep}) has two simple unit eigenvalues. For simplicity, we assume all other eigenvalues of $\F_d$ are zero. Under this assumption, the Jordan canonical form of $\F_d$ can be represented as:

\begin{equation}
\label{Fd TwoSimpleUnit Rep}
\begin{aligned}
\F_d&=XJX^{-1}\\
&=
\begin{pmatrix}
\co{1}{1}&\co{2}{1}&0&\dots&0&0\\
\co{1}{2}&\co{2}{2}&0&\dots&0&0\\
\co{1}{3}&\co{2}{3}&1&\dots&0&0\\
\vdots&\vdots&\vdots&\ddots&\vdots&\vdots\\
\co{1}{d-1}&\co{2}{d-1}&0&\dots&1&0\\
\co{1}{d}&\co{2}{d}&0&\dots&0&1\\
\end{pmatrix}
\begin{pmatrix}
1&&&&&\\
&1&&&&\\
&&0&&&\\
&&&\ddots&\\
&&&&0&\\
&&&&&0\\
\end{pmatrix}
\begin{pmatrix}
\co{1}{1}&\co{2}{1}&0&\dots&0&0\\
\co{1}{2}&\co{2}{2}&0&\dots&0&0\\
\co{1}{3}&\co{2}{3}&1&\dots&0&0\\
\vdots&\vdots&\vdots&\ddots&\vdots&\vdots\\
\co{1}{d-1}&\co{2}{d-1}&0&\dots&1&0\\
\co{1}{d}&\co{2}{d}&0&\dots&0&1\\
\end{pmatrix}^{-1}\\
&=
\begin{pmatrix}
1&0&0&\dots&0&0\\
0&1&0&\dots&0&0\\
-\frac{\Delta_{23}}{\Delta_{12}}&\frac{\Delta_{13}}{\Delta_{12}}&0&\dots&0&0\\
\vdots&\vdots&\vdots&\ddots&\vdots&\vdots\\
-\frac{\Delta_{2d-1}}{\Delta_{12}}&\frac{\Delta_{1d-1}}{\Delta_{12}}&0&\dots&0&0\\
-\frac{\Delta_{2d}}{\Delta_{12}}&\frac{\Delta_{1d}}{\Delta_{12}}&0&\dots&0&0\\
\end{pmatrix},
\end{aligned}
\end{equation}
where $\Delta_{ij}=\co{1}{i}\co{2}{j}-\co{1}{j}\co{2}{i}$. Obviously, the largest dimension of unit Jordan block in Eq (\ref{Fd TwoSimpleUnit Rep}) is one, and therefore, the process $\left\{X_t\right\}$ generated by this $\F_d$ is periodically integrated of order one, according to Definition \ref{Def:Periodic integration}. In addition, it turns out that the diagonalisable multi-companion matrix with two eigenvalues equal to one has a special form, in which the top left block is $2 \times 2$ diagonal matrix of ones. 

In conclusion, when $\F_d$ is shown in Eq (\ref{Fd TwoSimpleUnit Rep}), the generated series $\left\{X_t\right\}$ is periodically integrated of order one. Consequently, the model in Eq (\ref{PARp MCRep}) is a $\PIAR{1}{2}$ model. Following this conclusion, we are interested in finding a PI-filter which removes the two simple unit roots in the model and transforms $X_t$ from PI(1) to periodically stationary.

To specify the parameters of the PI-filter, we first consider the role of Jordan matrix. From Eq (\ref{ZT define}), when Jordan matrix has the form as shown in Eq (\ref{Fd TwoSimpleUnit Rep}), it indicates:

\begin{equation}
\label{ZT twosimple}
Z_T^{(1)}=Z_{T-1}^{(1)}+W_{T}^{(1)}, \quad
Z_T^{(2)}=Z_{T-1}^{(2)}+W_{T}^{(2)},
\end{equation}
where $Z_T^{(1)}$ and $Z_T^{(2)}$ are two random walks. Notice that the remaining processes of $\Z_T$, namely $Z_T^{(i)}$ for $i=3,\dots,d$, are all stationary.

Subsequently, we consider the role of similarity matrix, particularly focusing on the role of seed-parameters, see Eq (\ref{Role of similarity}). Expanding Eq (\ref{Role of similarity}) gives $X_{[T,s]}=\co{1}{d-s+1}Z_T^{(1)}+\co{2}{d-s+1}Z_T^{(2)}+\sum_{i=3}^{m}\co{i}{d-s+1}Z_T^{(i)}$, which indicates the two random walks together drive each seasonal component non-stationary, such that $\left\{X_{[T,s]}, T=1,2,\dots\right\} \sim \text{I}(1)$. In order to remove these two random walks from $\left\{X_t\right\}$ and transform $\left\{X_t\right\}$ to be periodically stationary, a PI-filter $(1-\theta_{1,s}L-\theta_{2,s}L^2)$ is introduced, where $\theta_{1,s}$ and $\theta_{2,s}$ for $s=1,\dots, d$ are determined by:

\begin{equation}
\label{PIparam TwoUnit PI1}
\begin{aligned}
&\theta_{1,s}=
\frac{\Delta_{d-s+1 \ d-s+3}}{\Delta_{d-s+2 \ d-s+3}}
=\frac{\co{1}{d-s+1}\co{2}{d-s+3}-\co{1}{d-s+3}\co{2}{d-s+1}}{\co{1}{d-s+2}\co{2}{d-s+3}-\co{1}{d-s+3}\co{2}{d-s+2}},\\
&\theta_{2,s}=
\frac{\Delta_{d-s+2 \ d-s+1}}{\Delta_{d-s+2 \ d-s+3}}
=\frac{\co{1}{d-s+2}\co{2}{d-s+1}-\co{1}{d-s+1}\co{2}{d-s+2}}{\co{1}{d-s+2}\co{2}{d-s+3}-\co{1}{d-s+3}\co{2}{d-s+2}}
,
\end{aligned}
\end{equation}
with

\begin{equation}
\label{COparam TwoUnit PI1}
\begin{aligned}
&\co{1}{d+k}=\co{1}{k}, &\co{2}{d+k}=\co{2}{k}, \quad k=1,2.
\end{aligned}
\end{equation}

Additionally, we find that this second order PI-filter is equivalent to a cascaded filter $(1-\beta_sL)(1-\alpha_sL)$ where

\begin{equation}
\label{PIparam TwoUnit PI1 cascade}
\begin{aligned}
&\alpha_s=\frac{c_{i_1}^{(d-s+1)}}{c_{i_1}^{(d-s+2)}},
&\beta_s=
\frac{\co{i_2}{d-s+1}-\alpha_s\co{i_2}{d-s+2}}{\co{i_2}{d-s+2}-\alpha_{s-1}\co{i_2}{d-s+3}},
\quad
s=1,\dots,d,
\end{aligned}
\end{equation}
with $(i_1,i_2)=(1,2)$ or $(i_1,i_2)=(2,1)$. Note that $\alpha_s$ and $\beta_s$ defined in Eq (\ref{PIparam TwoUnit PI1 cascade}) satisfy the restriction $\prod_{s=1}^{d}\alpha_s=\prod_{s=1}^{d}\beta_s=1$, and therefore, $(1-\beta_sL)$ and $(1-\alpha_sL)$ are two unit PI-filters. In addition, Eq (\ref{PIparam TwoUnit PI1 cascade}) shows there are two solutions for $\alpha_s$ and $\beta_s$ parameters, it is because the two random walks in Eq (\ref{ZT twosimple}) have a same integration order of one, and either of them can be firstly eliminated when applying $(1-\alpha_sL)$ to $\left\{X_t\right\}$. For instance, the solution of $\alpha_s$ and $\beta_s$ obtained by setting $(i_1,i_2)=(1,2)$ means $Z_T^{(1)}$ is firstly eliminated when applying $(1-\alpha_sL)$ to $\left\{X_t\right\}$, leaving $Z_T^{(2)}$ as the only non-stationary part which is then eliminated by applying $(1-\beta_sL)$ to $(1-\alpha_sL)X_t$. Furthermore, it can be proved that the two solutions of $\alpha_s$ and $\beta_s$ lead to a same result for PI-parameters $\theta_{i,s}$ shown in Eq (\ref{PIparam TwoUnit PI1}), such that $\theta_{1,s}=\alpha_s+\beta_s$ and $\theta_{2,s}=-\beta_s\alpha_{s-1}$.

\subsubsection{Two chained unit roots}

The second case arises when $\F_d$ in Eq (\ref{PARp MCRep}) has two chained unit eigenvalues. For simplicity, we assume all other eigenvalues of $\F_d$ are zero. Under this assumption, the Jordan canonical form of $\F_d$ is represented as:

\begin{equation}
\begin{aligned}
\label{Fd TwoChainedUnit Rep}
\F_d&=XJX^{-1}\\
&=
\begin{pmatrix}
\co{1}{1}&\co{2}{1}&0&\dots&0&0\\
\co{1}{2}&\co{2}{2}&0&\dots&0&0\\
\co{1}{3}&\co{2}{3}&1&\dots&0&0\\
\vdots&\vdots&\vdots&\ddots&\vdots&\vdots\\
\co{1}{d-1}&\co{2}{d-1}&0&\dots&1&0\\
\co{1}{d}&\co{2}{d}&0&\dots&0&1\\
\end{pmatrix}
\begin{pmatrix}
1&1&&&&\\
&1&&&&\\
&&0&&&\\
&&&\ddots&&\\
&&&&0&\\
&&&&&0\\
\end{pmatrix}
\begin{pmatrix}
\co{1}{1}&\co{2}{1}&0&\dots&0&0\\
\co{1}{2}&\co{2}{2}&0&\dots&0&0\\
\co{1}{3}&\co{2}{3}&1&\dots&0&0\\
\vdots&\vdots&\vdots&\ddots&\vdots&\vdots\\
\co{1}{d-1}&\co{2}{d-1}&0&\dots&1&0\\
\co{1}{d}&\co{2}{d}&0&\dots&0&1\\
\end{pmatrix}^{-1}\\
&=
\begin{pmatrix}
-\frac{\Delta_{21}+\co{1}{2}\co{1}{1}}{\Delta_{12}}&\frac{{(\co{1}{1})}^2}{\Delta_{12}}&0&\dots&0&0\\
-\frac{(\co{1}{2})^2}{\Delta_{12}}&\frac{\Delta_{12}+\co{1}{1}\co{1}{2}}{\Delta_{12}}&0&\dots&0&0\\
-\frac{\Delta_{23}+\co{1}{2}\co{1}{3}}{\Delta_{12}}&\frac{\Delta_{13}+\co{1}{1}\co{1}{3}}{\Delta_{12}}&0&\dots&0&0\\
\vdots&\vdots&\vdots&\ddots&\vdots&\vdots\\
-\frac{\Delta_{2d-1}+\co{1}{2}\co{1}{d-1}}{\Delta_{12}}&\frac{\Delta_{1d-1}+\co{1}{1}\co{1}{d-1}}{\Delta_{12}}&0&\dots&0&0\\
-\frac{\Delta_{2d}+\co{1}{2}\co{1}{d}}{\Delta_{12}}&\frac{\Delta_{1d}+\co{1}{1}\co{1}{d}}{\Delta_{12}}&0&\dots&0&0
\end{pmatrix}.
\end{aligned}
\end{equation}

Obviously, the largest dimension of unit Jordan block in Eq (\ref{Fd TwoChainedUnit Rep}) is two, and therefore, the process $\left\{X_t\right\}$ generated by this $\F_d$ is periodically integrated of order two, according to Definition \ref{Def:Periodic integration}. Correspondingly, the model in Eq (\ref{PARp MCRep}) is a $\PIAR{2}{2}$ model. Moreover, compared with the representation in Eq (\ref{Fd TwoSimpleUnit Rep}) where $\F_d$ has two simple unit eigenvalues, the $(1,2)$-th and $(2,1)$-th elements of $\F_d$ in Eq (\ref{Fd TwoChainedUnit Rep}) cannot be both equal to zero at the same time, since that will make the similarity matrix $X$ singular. Thus, the $2 \times 2$ upper-left corner of $\F_d$ is sufficient to distinguish the two cases where $\F_d$ has two simple or two chained unit eigenvalues.

Next, we derive a PI-filter which is utilized to transform $X_t \sim \text{PI}(2)$ into $X_t \sim \text{PI}(0)$. From Eq (\ref{ZT define}), the Jordan matrix with two chained unit eigenvalues indicates

\begin{equation}
Z_T^{(1)}=Z_{T-1}^{(1)}+Z_{T-1}^{(2)}+W_{T}^{(1)}, \quad
Z_T^{(2)}=Z_{T-1}^{(2)}+W_{T}^{(2)},
\end{equation}
where $Z_T^{(1)} \sim \text{I}(2)$ and $Z_T^{(1)} \sim \text{I}(1)$. It is worthwhile to mention that different from the previous situation with two simple unit eigenvalues, $Z_T^{(1)}$ process in this case has higher integration order due to the chained unit eigenvalues. In order to remove these two non-stationary parts, a PI-filter $(1-\theta_{1,s}L-\theta_{2,s}L^2)$ is introduced where $\theta_{1,s}$ and $\theta_{2,s}$ have a same general representation as shown in Eq (\ref{PIparam TwoUnit PI1}) but with

\begin{equation}
\label{COparam TwoUnit PI2}
\begin{aligned}
&\co{1}{d+k}=\co{1}{k}, &\co{2}{d+k}=\co{2}{k}-\co{1}{k}, \quad k=1,2.
\end{aligned}
\end{equation}

In this situation, the second order PI-filter $(1-\theta_{1,s}L-\theta_{2,s}L^2)$ is equivalent to a cascaded filter $(1-\beta_sL)(1-\alpha_sL)$ where $\alpha_s$ and $\beta_s$ are uniquely determined by 

\begin{equation}
\label{PIparam TwoUnit PI2 cascade}
\begin{aligned}
&\alpha_s=\frac{c_{1}^{(d-s+1)}}{c_{1}^{(d-s+2)}},
&\beta_s=
\frac{\co{2}{d-s+1}-\alpha_s\co{2}{d-s+2}}{\co{2}{d-s+2}-\alpha_{s-1}\co{2}{d-s+3}},
\quad
s=1,\dots,d.
\end{aligned}
\end{equation}

It is noted that the parameters $\alpha_s$ and $\beta_s$ in Eq (\ref{PIparam TwoUnit PI2 cascade}) also satisfy the restriction $\prod_{s=1}^{d}\alpha_s=\prod_{s=1}^{d}\beta_s=1$. Compared with the solutions of $\alpha_s$ and $\beta_s$ under two simple unit roots case, see Eq (\ref{PIparam TwoUnit PI1 cascade}), the solutions under two chained unit roots case are uniquely determined by the eigen information of the multi-companion matrix. It is because when there are two chained unit roots, the unit periodic filter $(1-\alpha_sL)$ is firstly applied to break the Jordan chain, and eliminate $Z_T^{(1)}$ process which has the highest integration order of two from $\left\{X_t\right\}$. After that, the only non-stationary part remaining in the process is $Z_T^{(2)}$ with integration order one, which makes the transformed series $(1-\alpha_sL)X_t$ periodically integrated of order one. Hereby, the unit periodic filter $(1-\beta_sL)$ is then applied to eliminate $Z_T^{(2)}$ from the transformed series and ensures $(1-\beta_sL)(1-\alpha_sL)X_t$ periodically stationary.

In conclusion, when $\F_d$ in Eq (\ref{PARp MCRep}) has two unit eigenvalues and all other eigenvalues have moduli strictly less than one, model Eq (\ref{PARp Rep}) can be rewritten as

\begin{equation*}
\begin{aligned}
&\psi_{p-2,s}(L)(1-\theta_{1,s}L-\theta_{2,s}L^2)X_t
=
\psi_{p-2,s}(L)(1-\beta_sL)(1-\alpha_sL)X_t
=
\eps_t, 
&t=1,2,\dots,
\end{aligned}
\end{equation*}
where $\psi_{p-2,s}(L)$ is a periodic autoregressive filter with order $p-2$. The second order PI-filter $(1-\theta_{1,s}L-\theta_{2,s}L^2)$ is determined by Eq (\ref{PIparam TwoUnit PI1}), where Eq (\ref{COparam TwoUnit PI1}) holds when there are two simple unit roots and Eq (\ref{COparam TwoUnit PI2}) holds when there are two chained unit roots. Equivalently, a cascade of two unit PI-filters $(1-\beta_sL)(1-\alpha_sL)$ can also be applied to transform $X_t$ into periodically stationary, which is determined by Eq (\ref{PIparam TwoUnit PI1 cascade}) and (\ref{PIparam TwoUnit PI2 cascade}) for two simple and two chained unit roots respectively.

\subsection{Multiple unit roots}

Based on previous two subsections, we extend the above conclusions to general cases. Consider a PIAR$(p)$ model in Eq (\ref{PARp Rep}) which has $m_1$ unit roots with $m_1 \in [1,p]$. Subsequently, Eq (\ref{PARp Rep}) can be rewritten as

\begin{equation}
\label{PIAR MultiStep}
\psi_{p-m_1,s}(L)(1-\theta_{1,s}L-\dots-\theta_{m_1,s}L^{m_1})X_t=\eps_t, 
\quad 
t=1,2,\dots,
\end{equation}
where $\psi_{p-m_1,s}(L)$ is a periodic autoregressive filter with order $p-m_1$ and particularly $\psi_{0,s}(L)=1$, and $(1-\theta_{1,s}L-\dots-\theta_{m_1,s}L^{m_1})$ is a PI-filter with order $m_1$. This PI-filter is used to eliminate all the $m_1$ unit roots from $\left\{X_t\right\}$ and transform $\left\{X_t\right\}$ into periodically stationary, such that $(1-\theta_{1,s}L-\cdots-\theta_{m_1,s}L^{m_1})X_t \sim \text{PI}(0)$. 

Moreover, similarly to previous two subsections, the PI-parameters in Eq (\ref{PIAR MultiStep}) can also be uniquely determined by the eigen information of the corresponding multi-companion matrix $\F_d$ in Eq (\ref{PARp MCRep}). Here, we provide a general parametrization result for PI-parameters.

Let $\theta(s)=(\theta_{1,s},\dots,\theta_{m_1,s})^{'}$ be the PI-parameters at season $s$. We construct a $d \times m_1$ matrix $X^{(1)}$, such that

\begin{equation}
\label{X1 Define}
X^{(1)}
=
\begin{pmatrix}
\co{1}{1}&\co{2}{1}&\dots&\co{m_1}{1}\\
\co{1}{2}&\co{2}{2}&\dots&\co{m_1}{2}\\
\vdots&\vdots&\ddots&\vdots\\
\co{1}{d}&\co{2}{d}&\dots&\co{m_1}{d}
\end{pmatrix},
\end{equation}
which is the top-left part of the similarity matrix $X$ of $\F_d$. Note that $X^{(1)}$ collects all the seed-vectors corresponding to the $m_1$ unit eigenvalues of $\F_d$. Given the special property of the multi-companion matrix, see \citet[][after Eq 5.4]{boshnakov2002multi}, $X^{(1)}$ in Eq (\ref{X1 Define}) is sufficient to determine the entire information of the eigenvectors associated with the $m_1$ unit eigenvalues of $\F_d$. This property is useful when estimating the eigenvectors associated with the $m_1$ unit eigenvalues of $\F_d$. In particular, when $m>d$, this property helps to reduce the number of unknowns from $mm_1$ to $dm_1$. 

After that, an $m_1 \times 2d$ matrix $\Xbind$ is created as:

\begin{equation}
\label{Xbind Define}
\Xbind
=
\begin{pmatrix}
X^{(1)}\Junit\\
X^{(1)}
\end{pmatrix}^{'},
\end{equation}
where $\Junit$ is an $m_1 \times m_1$ unit Jordan matrix defined by Eq (\ref{Junit}). We find that the PI-parameters $\theta(s)$ in Eq (\ref{PIAR MultiStep}) are uniquely determined by solving

\begin{equation}
\label{PIparam determination}
\begin{aligned}
&(\Xbind)_{\bullet (d-s+2):(d-s+m_1+1)}\theta(s)=(\Xbind)_{\bullet d-s+1},
&s=1,\dots,d,
\end{aligned}
\end{equation}
where $(\Xbind)_{\bullet j}$ stands for the $j$-th column of $\Xbind$. The uniqueness of $\theta(s)$ is guaranteed by the linear independence of columns of $X^{(1)}$. 

Based on the above results, we propose a new estimation method for PIAR$(p)$ models which uses the eigen information of the multi-companion matrix in their multi-companion representations. A special case happens when a PIAR$(p)$ model has exactly $p$ unit roots. In this case, Eq (\ref{PIAR MultiStep}) reduces to $(1-\theta_{1,s}L-\dots-\theta_{p,s}L^p)X_t=\eps_t$, and the $\F_d$ matrix in its multi-companion representation has $p$ unit eigenvalues and all other eigenvalues are zero. Due to the special properties of the multi-companion matrix, see \citet[][Lemma1]{boshnakov2009generation} and \citet[][after Eq 5.4]{boshnakov2002multi}, the number of $dp$ seed-parameters which are collected in $X^{(1)}$ is sufficient to determine the entire eigen information of $\F_d$. In turn, the $dp$ seed-parameters are also sufficient to determine the PI-parameters $\theta(s)$ for all seasons by using Eq (\ref{PIparam determination}). Therefore, we regard Eq  (\ref{PIparam determination}) as a bridge to transfer the eigen information of $\F_d$ into the information of the parameters of the PI-filter. Moreover, instead of estimating the PI-parameters directly, we estimate the eigen information of $\F_d$, or more precisely, the seed-parameters of $\F_d$. At last, an optimization routine is applied to find the estimators of the seed-parameters which minimize the residual sum of squares of the PIAR$(p)$ model, and the estimated PI-parameters can be obtained by solving Eq (\ref{PIparam determination}). 

A more general case happens when a PIAR$(p)$ model has $m_1$ unit roots where $p>m_1$. In this case, Eq (\ref{PIAR MultiStep}) can be viewed as a two-step process such that 

\begin{equation}
\label{PIAR twostep}
\left\{
\begin{aligned}
&(1-\theta_{1,s}L-\cdots-\theta_{m_1,s}L^{m_1})X_t=y_t, \\
&(1-\psi_{1,s}L-\cdots-\psi_{p-m_1,s}L^{p-m_1})y_t=\eps_t,\\
\end{aligned}
\right.
\end{equation}
where the first and the second are PIAR$(m_1)$ and PAR$(p-m_1)$ processes respectively. It is worth noting that the parameters of these two steps in Eq (\ref{PIAR twostep}) can be estimated separately. In the first step, given that the PIAR$(m_1)$ process has exactly $m_1$ unit roots, we can construct $X^{(1)}$ and $\Xbind$, and the estimators of PI-parameters $\theta(s)$ are obtained by solving Eq (\ref{PIparam determination}). After that, applying the PI-filter to $X_t$ transforms $X_t$ into periodically stationary, and therefore, the second step is a PAR$(p-m_1)$ process which can either be estimated by periodic Yule-Walker \citep[see][]{pagano1978periodic} or weighted least squares \citep[see][]{basawa2001large}.

A key point to emphasize is that rather than estimating PI-parameters of a PIAR$(p)$ model directly, our estimation method sets the seed-parameters of the multi-companion matrix as the unknowns. Subsequently, Eq (\ref{PIparam determination}) is utilized as a bridge to transfer the estimation information of seed-parameters to PI-parameters. As a result, our method offers a significant advantage over the existing method \cite{boswijk1997multiple} that requires dealing with non-linear restrictions between PI-parameters, and also extends the current literature which mainly deals with quarterly PIAR models to general cases. In particular, it is found that by setting $d=4$, the quarterly PI-parameters derived by solving Eq (\ref{PIparam determination}) automatically satisfy the non-linear restrictions given by \cite{boswijk1997multiple} for a single, two and three unit roots cases. Hereby, the approach used in \cite{boswijk1997multiple} is a special case of our multi-companion method.



\section{Monte Carlo Analysis}
\label{Monte Carlo Analysis}

This section provides the results of Monte Carlo experiments to verify the estimation method of periodically integrated autoregressive models, with the theoretical analysis introduced in Section \ref{Multi-companion method applied to PIAR model}.

Firstly, in order to generate the periodically integrated series, we use the method introduced by \cite{boshnakov2009generation} which is based on the multi-companion representation in Eq (\ref{PARp MCRep}) and the eigen information of the multi-companion matrix. 

Table \ref{tab:EigInf generatePI} provides the eigen information of the multi-companion matrices, which is used to generate the periodically integrated series with quarterly period. The notation $c_i$ represents the $i$-th eigenvector (or seed-vector) associated with the $i$-th unit eigenvalue of the multi-companion matrix, and $\co{i}{j}$ means the $j$-th element of the $i$-th eigenvector. Note that all the unit eigenvalues given in Table \ref{tab:EigInf generatePI} are simple, which results in the generated series having periodic integration order one. Moreover, the remaining eigenvalues of the multi-companion matrices from Model I to Model III are zeros. However, Table \ref{tab:EigInf generatePI} does not include the information for eigenvectors corresponding to zero eigenvalues, since they are just standard basis with appropriate arrangement \citep[see][Lemma 1]{boshnakov2009generation}. 

\begin{table}[H]
	\centering
	\caption{Eigen information of the multi-companion matrix used to generate quarterly periodically integrated series}
	\label{tab:EigInf generatePI}
	\renewcommand{\arraystretch}{0.65}
	\begin{tabularx}{\textwidth}{XXXXX}
		\toprule
		\multicolumn{5}{c}{Model I: 
			$\PIAR{1}{1}$}\\
		& $j=1$ & $j=2$ & $j=3$ & $j=4$\\
	    $\lambda_1=1;\co{1}{j}$&-0.64&0.46&0.65&0.68\\
		\bottomrule
	\end{tabularx}
	\begin{tabularx}{\textwidth}{XXXXX}
		\toprule
		\multicolumn{5}{c}{Model II: 
			$\PIAR{1}{2}$}\\
			& $j=1$ & $j=2$ & $j=3$ & $j=4$\\
		$\lambda_1=1;\co{1}{j}$&0.08&-0.41&0.52&0.40\\
		$\lambda_2=1;\co{2}{j}$&0.22&0.29&-0.58&-0.49\\
		\bottomrule
	\end{tabularx}
	\begin{tabularx}{\textwidth}{XXXXX}
		\toprule
		\multicolumn{5}{c}{Model III: 
			$\PIAR{1}{3}$}\\
		& $j=1$ & $j=2$ & $j=3$ & $j=4$\\
	$\lambda_1=1;\co{1}{j}$&-0.64&-0.46&0.65&0.68\\
	$\lambda_2=1;\co{2}{j}$&-0.23&0.95&-0.83&-0.89\\
	$\lambda_3=1;\co{3}{j}$&-0.30&0.91&0.47&-0.15\\
		\bottomrule
	\end{tabularx}
\end{table}

The models in Table~\ref{tab:EigInf generatePI} are with one, two and three simple unit
roots, respectively. The corresponding periodic filter representations of the models are:

\begin{itemize}
	\item Model I: $X_t=\theta_{1,s}X_{t-1}+\eps_t$ where $\eps_t \sim N(0,\sigma_s^2)$;
	\item Model II: $X_t=\theta_{1,s}X_{t-1}+\theta_{2,s}X_{t-2}+\eps_t$ where $\eps_t \sim N(0,\sigma_s^2)$;
	\item Model III: $X_t=\theta_{1,s}X_{t-1}+\theta_{2,s}X_{t-2}+\theta_{3,s}X_{t-3}+\eps_t$ where $\eps_t \sim N(0,\sigma_s^2)$.
\end{itemize}

The numerical values of the PI-parameters $\theta_{i,s}$ and the variance of periodic white noise $\sigma_s^2$, are listed in Table
\ref{tab:SimulationResults} in the rows designated as `true' values.

The simulation starts by setting the sample size of the generated series as 240 and the simulation for each model runs 2000 times. Table \ref{tab:SimulationResults} shows the mean, standard deviation (sd) and root mean squared error (RMSE) of the estimated parameters. We observe that across all three models, the mean values of the estimated parameters derived from 2000 simulations closely align with the true values. Furthermore, the standard deviations and RMSE values are relatively low, indicating the robustness of our estimation method.

\begin{table}[H]
	\caption{Simulation results: mean value, standard deviation (sd) and root mean squared error (RMSE) of parameter estimates from simulation for Model I to Model 	III}
	\label{tab:SimulationResults}
	\begin{threeparttable}
		\begin{tabular}[t]{ccccccccccccc}
			\toprule
			Model I:& $\theta_{1,1}$ & $\theta_{1,2}$ & $\theta_{1,3}$ & $\theta_{1,4}$ & $\sigma_1^2$ & $\sigma_2^2$ & $\sigma_3^2$ & $\sigma_4^2$ &&&& \\
			\midrule
			true& -1.07 & 0.95 & 0.70 & -1.41 & 0.15 & 0.46 & 0.24 & 0.08 &&&&\\
			mean&  -1.07 & 0.95 & 0.70 & -1.41 & 0.15 & 0.45 & 0.23 & 0.07&&&& \\
			sd& 0.01 & 0.02 & 0.01 & 0.01 & 0.02 & 0.07 & 0.04 & 0.01&&&&\\
			RMSE& 0.01 & 0.02 & 0.01 & 0.01 & 0.02 & 0.07 & 0.04 & 0.01&&&&\\
			\bottomrule
			\toprule
			Model II & $\theta_{1,1}$ & $\theta_{1,2}$ & $\theta_{1,3}$ & $\theta_{1,4}$ & $\theta_{2,1}$ & $\theta_{2,2}$ & $\theta_{2,3}$ & $\theta_{2,4}$ &$\sigma_1^2$ & $\sigma_2^2$ & $\sigma_3^2$ & $\sigma_4^2$ \\
			\midrule
			true&  -0.73 & 1.26 & -4.00 & -1.85 & -1.12 & 0.16 & 4.17 & -1.31 & 0.29 & 0.37 & 0.44 & 0.02 \\
			mean& -0.72 & 1.27 & -4.00 & -1.86 & -1.10 & 0.16 & 4.15 & -1.33 & 0.28 & 0.37 & 0.43 & 0.02 \\
			sd& 0.02 & 0.02 & 0.05 & 0.01 & 0.02 &  $<0.01$& 0.08 & 0.03 & 0.05 & 0.07 & 0.08 & $<0.01$\\
			RMSE&0.02 & 0.02 & 0.05 & 0.01 & 0.03 &  $<0.01$& 0.08 & 0.04 & 0.05 & 0.07 & 0.08 &  $<0.01$\\
			\bottomrule
			\toprule
			Model III &$\theta_{1,1}$ & $\theta_{1,2}$ & $\theta_{1,3}$ & $\theta_{1,4}$ & $\theta_{2,1}$ & $\theta_{2,2}$ & $\theta_{2,3}$ & $\theta_{2,4}$ &$\theta_{3,1}$ & $\theta_{3,2}$ & $\theta_{3,3}$ & $\theta_{3,4}$\\
			\midrule
			true& -0.16 & 1.83 & 1.10 & -3.21 & -0.5 & 0.28 & -2.01 & 3.53 & 0.55 & 0.91 & -0.31 & -6.45 \\
			mean&  -0.15 & 1.83 & 1.10 & -3.23 & -0.5 & 0.28 & -2.02 & 3.56 & 0.55 & 0.91 & -0.31 & -6.52 \\
			sd& $<0.01$ & 0.01 & 0.01 & 0.03 & $<0.01$& $<0.01$ & 0.02 & 0.04 & $<0.01$ & 0.01 & $<0.01$& 0.08\\
			RMSE&$<0.01$& 0.01 & 0.01 & 0.03 & $<0.01$ & $<0.01$& 0.02 & 0.05 & $<0.01$ & 0.01 & $<0.01$& 0.10 \\
			\midrule
			&$\sigma_1^2$ & $\sigma_2^2$ & $\sigma_3^2$ & $\sigma_4^2$ &&&&&&&&\\
			\midrule
			true & 0.22 & 0.35 & 0.25 & 0.05&&&&&&&&\\\
			mean & 0.22 & 0.35 & 0.25 & 0.05&&&&&&&&\\
			sd &0.04 & 0.05 & 0.04 & 0.01&&&&&&&&\\
			RMSE &0.04 & 0.05 & 0.04 & 0.01&&&&&&&&\\
			\bottomrule
		\end{tabular}
	\end{threeparttable}
\end{table}

It is worthwhile to point out that, our method avoids considering the non-linear restrictions between PI-parameters for Model I to III during the estimation process. However, it can be checked that in each simulation from Model I to III,  the PI-parameters derived by our estimation method automatically satisfy the non-linear restrictions given in \cite{boswijk1997multiple}. This outcome provides additional affirmation of the effectiveness and validity of our estimation approach.


\section{Application}
\label{Application}

In this section, we apply periodically integrated autoregressive models to forecast future
values of U.S. monthly electricity end use, and compare the forecasting performance of the
PIAR model with a non-periodic model (namely ARIMA) and a PAR model. The data is downloaded
from Monthly Energy Review from U.S. Energy Information Administration \footnote{https://www.eia.gov/totalenergy/data/monthly/}. The series contains
50 years of data from January 1973 to November 2022, and they are measured in Billion
Kilowatt-hours (BKWh). We partition the series into two sets, one consisting of observations
from January 1973 to December 2019 (47 years) used for model estimation, and the other
containing out-of-sample data from January 2020 to November 2022 used for forecasting
comparison. As the series is recorded monthly, we assume the period of the series is $d=12$.
So, the sample size of observations used for model estimation is
$n=N \times d=47 \times 12=564$.

\begin{figure}[ht]
	\centering		
	\includegraphics[width=\linewidth, height=0.35\textheight, keepaspectratio]{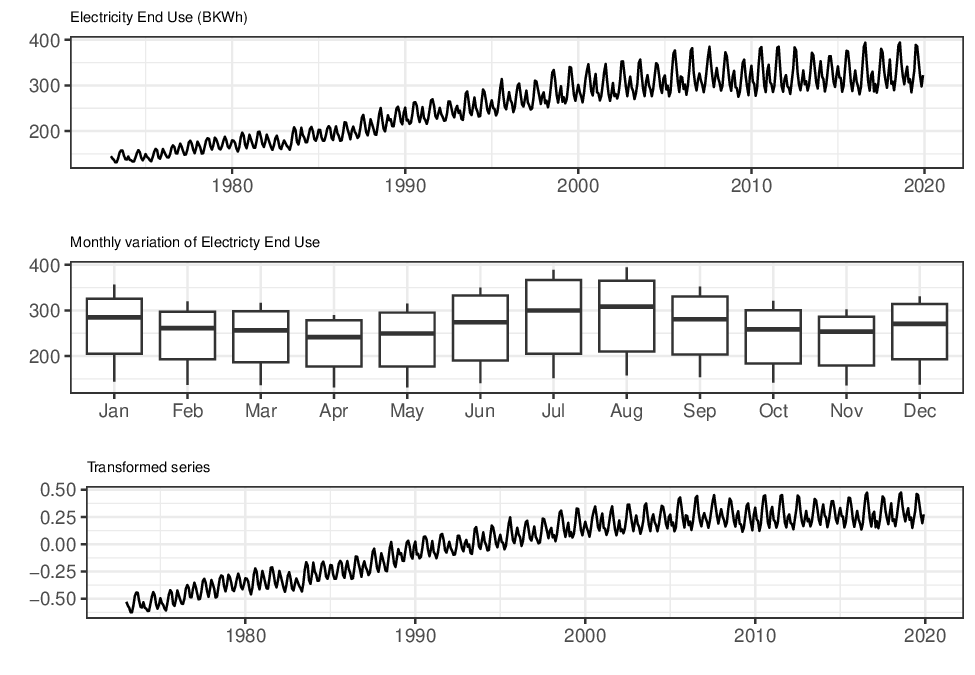}
	\caption{U.S. monthly electricity end use in BKWh (top), seasonal boxplots (middle),
          and the centred log-transformed series (bottom)}
	\label{Fig:ElecEndUseOverall}
\end{figure}

For $N=47$ years of data as shown in the top plot of Figure \ref{Fig:ElecEndUseOverall},
the series exhibits significant monthly variation and an upward trend. The monthly
variation is also seen from the middle plot of Figure \ref{Fig:ElecEndUseOverall}, where the electricity use remains relatively high both in summer (July and Aug) and in winter (Jan and Dec). Sometimes a log-transformation can remove the seasonal variation (by turning it into a seasonal mean or `level') but not here. Indeed, the bottom graph in Figure~\ref{Fig:ElecEndUseOverall} shows the log-transformed series, centred by subtracting the overall mean.

A PAR(5) model is firstly considered to fit the series, where the AIC and BIC reach their minimum values at -2697 and -2433 respectively. We find that the estimated multi-companion matrix of the PAR(5) model exhibits a pair of eigenvalues whose moduli approximate unity. This indicates the existence of two unit roots in the process, and hereby, a PIAR model with two simple and two chained unit roots should be both considered. As the number of unit roots in a periodic time series does not affect the autoregressive order selection \citep[see][]{boswijk1997multiple}, we can fix the order to be $p=5$ when fitting the PIAR models. Therefore, a $\PIAR{1}{5}$ model with two simple unit roots and a $\PIAR{2}{5}$ model with two chained unit roots are then constructed. 

To determine whether there are two simple or two chained unit roots in the process, a likelihood ratio test is performed. Two separated null hypotheses are set as: the process has two simple unit roots (i.e. $\PIAR{1}{5}$) and the process has two chained unit roots (i.e. $\PIAR{2}{5}$), respectively. The alternative hypothesis is the process does not include any unit roots (i.e. PAR(5)). The likelihood ratio test statistic is calculated by $Q_{LR}=N\log(|S^{-1}S_0|)$ where $S_0$ and $S$ are the residual sum of squares matrix under the null and alternative hypothesis respectively. Under the null, the test statistic should follow an asymptotic distribution \citep[see][Thm. 5.4.2]{YueyunZhuPARandPIARmodels} and the corresponding quantile values can be found from \citet[Table 15.1]{johansen1995likelihood}. The result turns to be that we accept the null that there are two chained unit roots in the process (with $Q_{LR}=4.08 < 12.21$). In addition, the $\PIAR{2}{5}$ model has smallest AIC and BIC values compared with $\PIAR{1}{5}$ and PAR(5). Therefore, we choose a $\PIAR{2}{5}$ as the final model. Let $\left\{X_t, t=1,\dots,n\right\}$ be the log-transformed series, and write the representation of the $\PIAR{2}{5}$ as

\begin{equation*}
(1-\psi_{1,s}L-\psi_{2,s}L^2-\psi_{3,s}L^3)(1-\beta_sL)(1-\alpha_sL)X_t=\eps_t,
\quad
\eps_t \sim N(0,\sigma_s^2),
\end{equation*}
where $\psi_{3,s}(L)=1-\psi_{1,s}L-\psi_{2,s}L^2-\psi_{3,s}L^3$ is the periodic autoregressive filter of order 3; $(1-\alpha_sL)$ and $(1-\beta_sL)$ are two unit PI-filters with $\prod_{s=1}^{d}\alpha_s=\prod_{s=1}^{d}\beta_s=1$. 

The two-step method introduced in Eq (\ref{PIAR twostep}) is applied to estimate the above $\PIAR{2}{5}$ model and the estimation result is in Table \ref{tab:ParamTablepi2ar5}. Table \ref{tab:ParamTablepi2ar5} shows the estimated parameters $\hat{\alpha}_s$ and $\hat{\beta}_s$ which satisfy the restrictions $\prod_{s=1}^{d}\alpha_s=\prod_{s=1}^{d}\beta_s=1$. Moreover, it can be proved that the roots of the set of polynomials $\left\{\hat{\psi}_{3,s}(L)\right\}_{s=1}^{d}$ are outside the unit circle, and hereby, the filter $\hat{\psi}_{3,s}(L)$ is a periodically autoregressive filter. 

\begin{table}[ht]
	\caption{Parameter estimates of $\PIAR{2}{5}$ model}
	\label{tab:ParamTablepi2ar5}
	\begin{threeparttable}
		\begin{tabular}[t]{ccccccccccccc}
			\toprule
			season & s=1 & s=2 & s=3 & s=4 & s=5 & s=6 & s=7 & s=8 & s=9 & s=10 &
			s=11 & s=12 \\
			\midrule
			$\hat{\alpha}_{s}$ & 1.692 & -0.256 & 1.246 & 5.698 & 0.506 & 3.748 & 2.026 & 1.188 & -1.894
			& 2.501 & 1.447 & -0.855 \\
			$\hat{\beta}_{s}$ &-0.764 & 1.131 & -0.168 & -5.020 & 0.588 & -2.988 & -1.056 & -0.230 &
			2.695 & -1.360 & -0.474 & 1.852 \\
			$\hat{\psi}_{1,s}$ &-0.749 & 1.065 & -0.830 & -5.136 & 0.352 & -2.707 & -0.316 & -0.184 &
			2.497 & -1.724 & -0.438 & 1.722 \\
			$\hat{\psi}_{2,s}$&1.308 & 0.440 & 0.397 & -1.278 & 1.376 & 0.787 & -0.756 & -0.100 & 0.325
			& 4.300 & -0.780 & 0.133 \\
			$\hat{\psi}_{3,s}$&0.163 & -0.526 & -0.055 & 0.899 & -0.172 & 3.362 & -0.027 & -0.230 &
			0.077 & 0.590 & 2.087 & -0.095 \\
			$\hat{\sigma}_s$ &0.021 & 0.023 & 0.016 & 0.016 & 0.019 & 0.021 & 0.025 & 0.020 & 0.017 &
			0.014 & 0.015 & 0.022 \\
			\bottomrule
		\end{tabular}
	\end{threeparttable}
\end{table}

\begin{figure}[ht]
	\centering		
	\includegraphics[width=\linewidth, height=0.35\textheight, keepaspectratio]{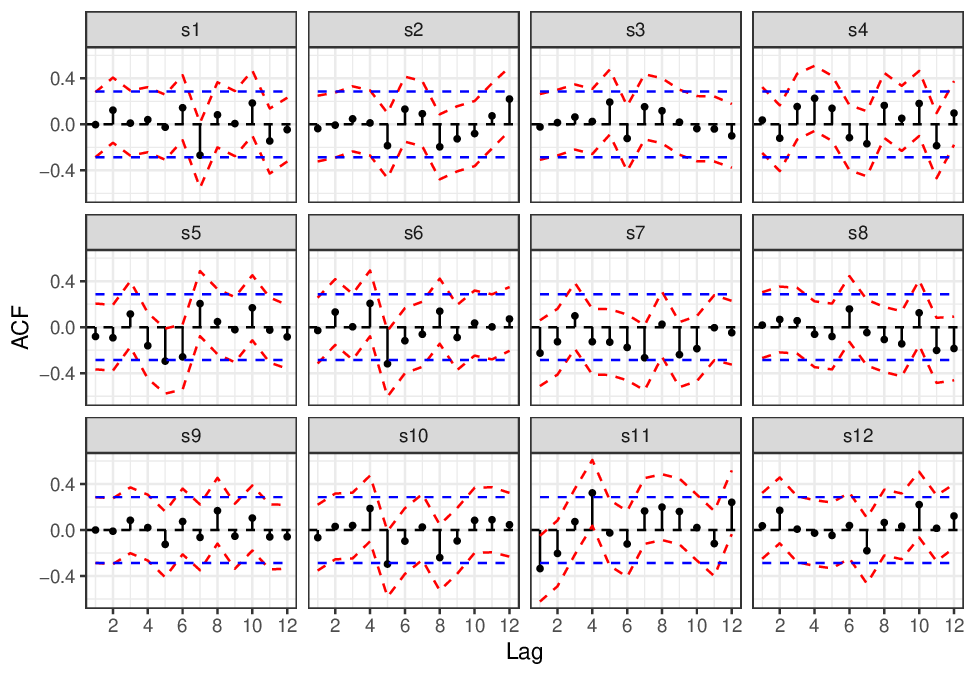}
	\caption{Residual periodic autocorrelations from $\PIAR{2}{5}$ model}
	\label{Fig:PeACFResTwoChained}
\end{figure}

Indeed, the $\PIAR{2}{5}$ model is found adequate to capture the periodically integrated
structure of transformed data of the U.S. monthly electricity end use. The adequacy is
visually validated by Figure \ref{Fig:PeACFResTwoChained}, where the periodic
autocorrelations of the residuals at each season are approximately located within the dashed
blue lines (namely $\pm 1.96/\sqrt{N}$). This suggests that the periodic autocorrelations of
residuals of the $\PIAR{2}{5}$ model are insignificant at each season. On the other hand, the
adequacy of model $\PIAR{2}{5}$ is numerically validated by Table
\ref{tab:McLeodTableForResiduals}, where the modified portmanteau McLeod test statistic
\citep[see][Eq 4.5]{mcleod1994diagnostic} is calculated by setting the maximum lag equal to
12. Table \ref{tab:McLeodTableForResiduals} shows that except for two seasons ($s=7$ and
$s=11$), the periodic autocorrelations of residuals for the other seasons at lag
$1,2,\dots,12$ are approximately equal to zero. Combined with the information delivered by
Figure \ref{Fig:PeACFResTwoChained}, we have the conclusion that the residuals of
$\PIAR{2}{5}$ model are periodically uncorrelated with each other.

\begin{table}[ht]
	\caption{McLeod portmanteau test statistic for residuals from $\PIAR{2}{5}$ model}
	\label{tab:McLeodTableForResiduals}
	\begin{threeparttable}
		\begin{tabular}[t]{ccccccccccccc}
			\toprule
			season & s=1 & s=2 & s=3 & s=4 & s=5 & s=6 & s=7 & s=8 & s=9 & s=10 &
			s=11 & s=12 \\
			\midrule
			McLeod stat& 8.39 & 8.65 & 5.21 & 12.38 & 13.81 & 10.08 & $14.53^{*}$ & 8.11 &
			3.87 & 10.50 & $21.03^{*}$ & 6.45 \\
			\bottomrule
		\end{tabular}
		\footnotesize{
			* An asterisk indicates the residual periodic autocorrelations at that season are significant, given that $\chi^2_{7}=14.07$ at 5\% level.}
	\end{threeparttable}
\end{table}

\begin{figure}[ht]
	\centering
	\begin{subfigure}[b]{0.45\textwidth}
		\centering
		\includegraphics[width=\linewidth]{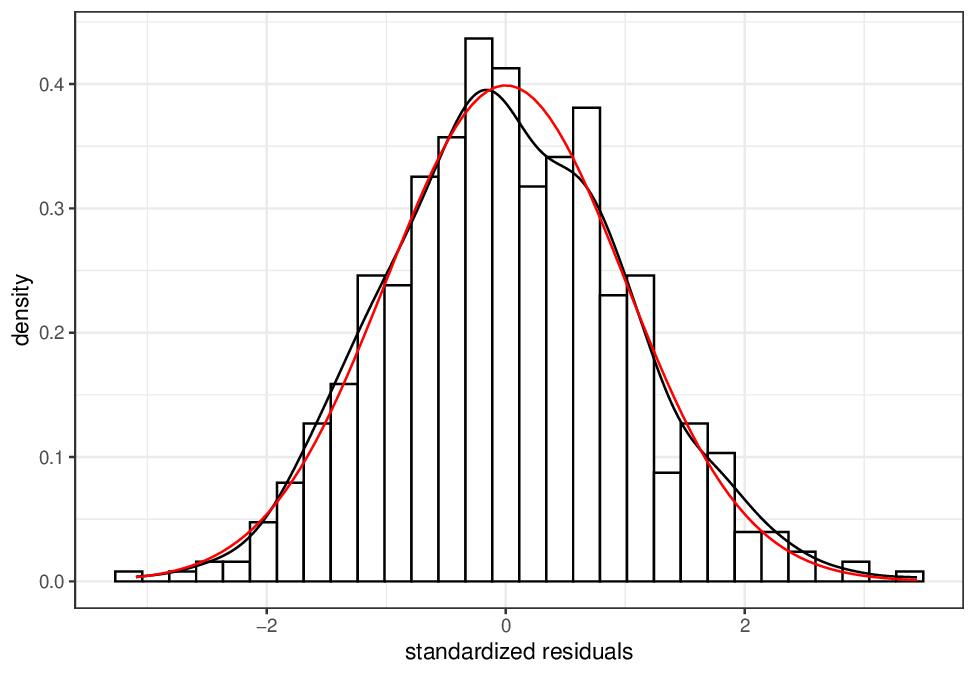}
		\caption{Density plot for standardized residuals}
		\label{Fig:DensityPlotStdResidual}
	\end{subfigure}
	\begin{subfigure}[b]{0.45\textwidth}
		\centering
		\includegraphics[width=\linewidth]{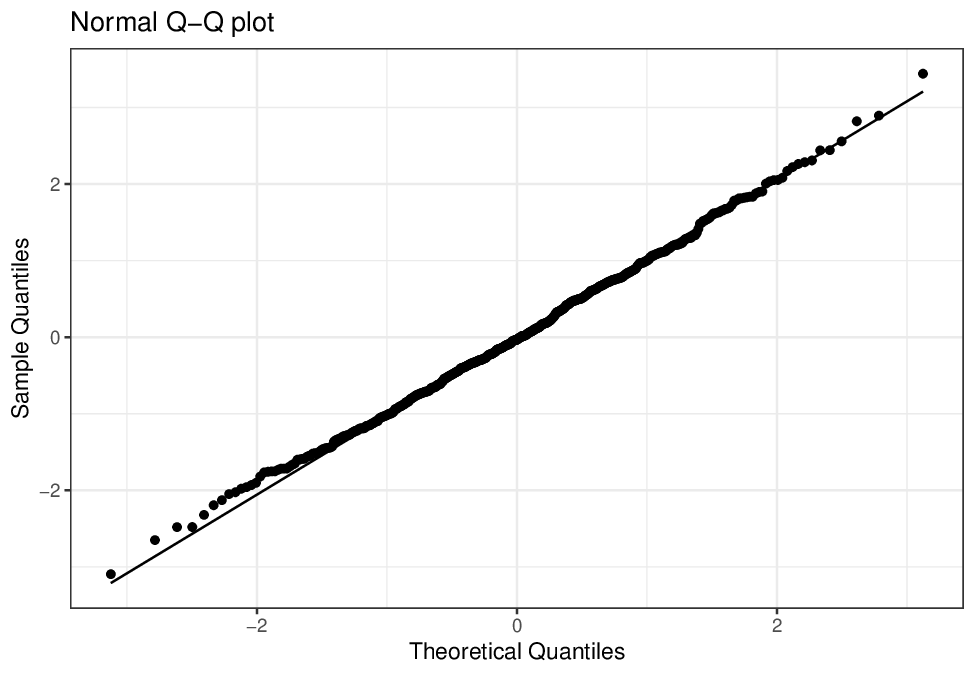}
		\caption{The quantile-quantile plot for standardized residuals}
		\label{Fig:QQPlotStdResidual}
	\end{subfigure}
	\caption{Normality check for standardized residuals of $\PIAR{2}{5}$ model}
	\label{Fig:StdResidual}
\end{figure}

Moreover, we check the normality of the standardized residuals. The standardized residuals are obtained from the original residuals of $\PIAR{2}{5}$ divided by their seasonal standard deviation namely $\hat{\sigma}_s$ given in Table \ref{tab:ParamTablepi2ar5}. Figure \ref{Fig:StdResidual} shows the density and the Q-Q plot of the standardized residuals, which indicates the standardized residuals are approximately normally distributed. In conclusion, the residuals of $\PIAR{2}{5}$ are periodic white noise and are normally distributed with mean 0 and variance $\hat{\sigma}_s^2$. Therefore, the $\PIAR{2}{5}$ model is verified to be adequate to capture the periodically integrated structure of the series.

\begin{figure}[ht]
	\centering
	\begin{subfigure}[b]{0.45\textwidth}
		\centering
		\includegraphics[width=\linewidth]{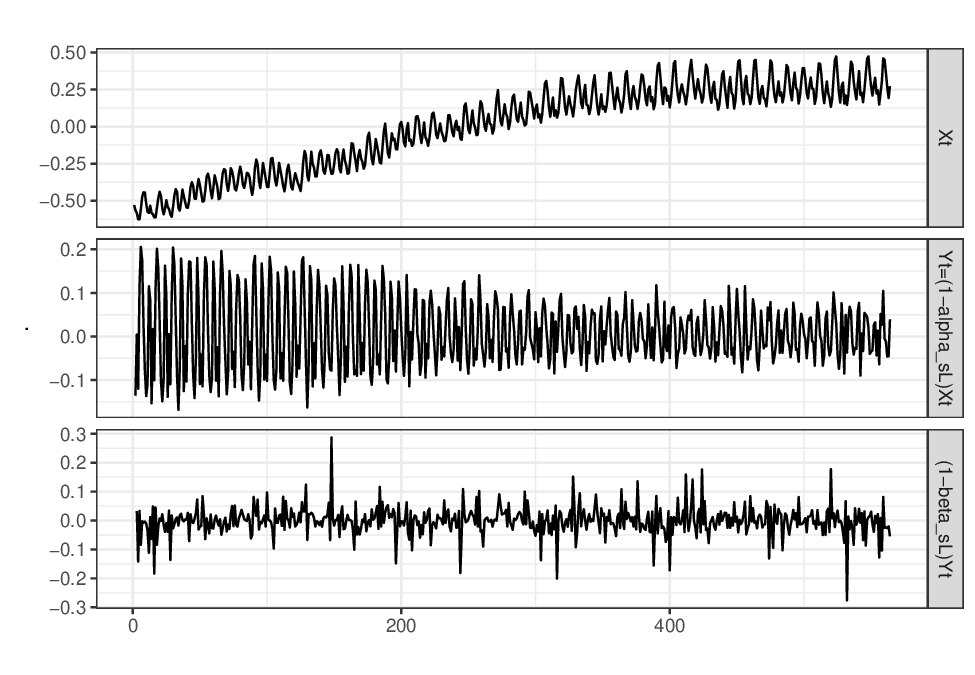}
		\caption{The effect of first-order and second-order PI-filters on univariate series}
		\label{Fig:PIdiffTSPlot}
	\end{subfigure}
	\begin{subfigure}[b]{0.45\textwidth}
		\centering
		\includegraphics[width=\linewidth]{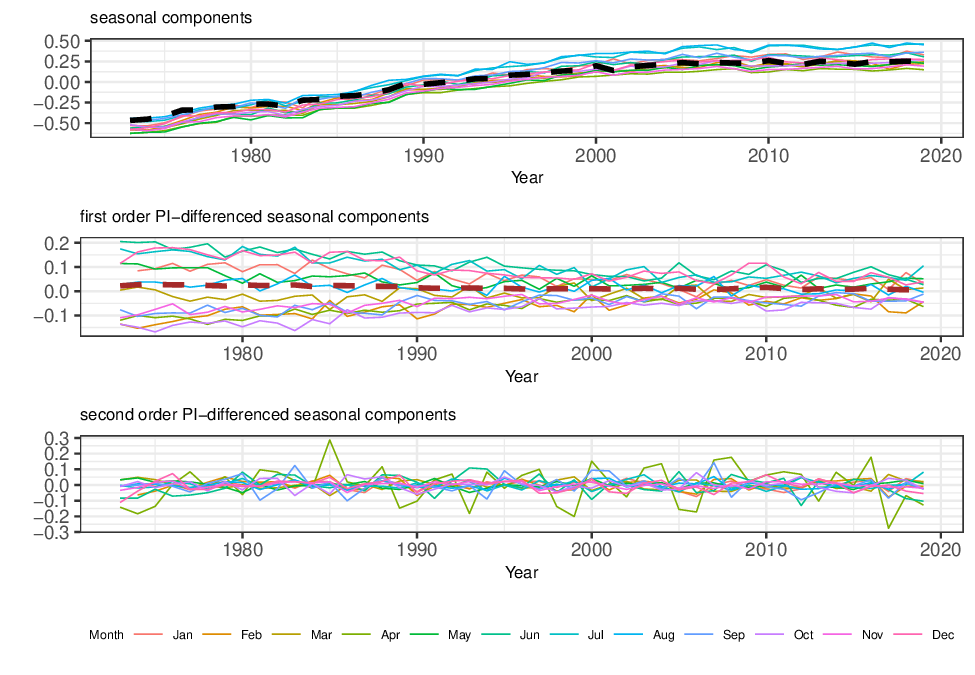}
		\caption{The effect of first-order and second-order PI-filters on seasonal components}
		\label{Fig:Bymonth}
	\end{subfigure}
	\caption{The effect of PI-filters on univariate series and its seasonal components. The black and brown dashed lines in \ref{Fig:Bymonth} represent the integrated series $Z_T^{(1)}$ and $Z_T^{(2)}$ respectively.}
	\label{Fig:PIexplanation}
\end{figure}

Next, we provide an explanation of the second order PI-filter $(1-\beta_sL)(1-\alpha_sL)$ in this $\PIAR{2}{5}$ model. Figure~ref{Fig:PIexplanation} visually illustrates the impact of the PI-filter on centred log-transformed series and their seasonal components. A transparent upward trend is observed in the top plot of Figure~\ref{Fig:PIdiffTSPlot}. This trend arises due to the dominant influence of the second-order integrated series $Z_T^{(1)}$ (black dashed line, as shown in the top plot of Figure~\ref{Fig:Bymonth}). This influence contributes to the emergence of a consistent upward pattern across each seasonal component. After applying the first-order PI-filter $(1-\alpha_sL)$ to $\left\{X_t\right\}$, it becomes apparent that the upward trend is eliminated from the series, as demonstrated in the middle plot of Figure~\ref{Fig:PIdiffTSPlot}. Correspondingly, the middle plot of Figure~\ref{Fig:Bymonth} highlights the impact of first-order integrated series $Z_T^{(2)}$ (brown dashed line), which forces the seasonal components drifting around $Z_T^{(2)}$. At last, the application of $(1-\beta_sL)(1-\alpha_sL)$ transforms $\left\{X_t\right\}$ into a periodically stationary series, see the bottom plot of Figure~\ref{Fig:PIdiffTSPlot}. Additionally, this second-order PI-filter eliminates the two integrated series which ensures the seasonal components stationary, as shown in the bottom plot of Figure~\ref{Fig:Bymonth}.

\begin{figure}[ht]
	\centering
	\includegraphics[width=\linewidth, height=0.35\textheight, keepaspectratio]{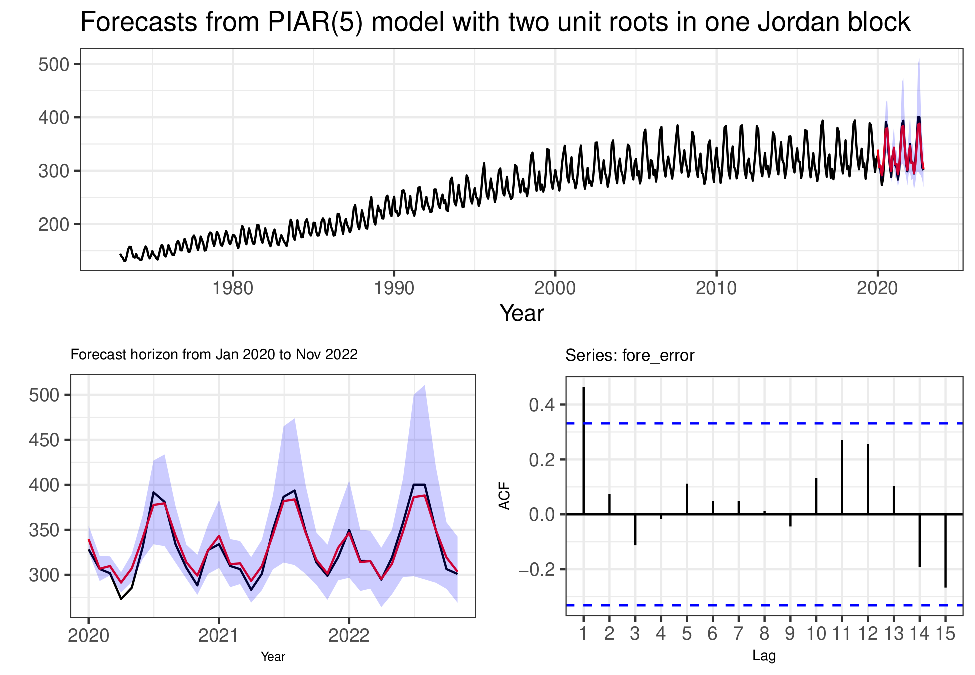}
	\caption{Forecast result from $\PIAR{2}{5}$ model, where the top shows actual data (black line) over the whole observation period with the forecasts (red line) and its confidence interval (blue ribbon) covering the pediciton period, the bottom-left panel displays a zoomed-in version for the forecasting period, the bottom-right panel shows the ACF plot for forecast error.}
	\label{Fig:ForecastPI2AR5}
\end{figure}

Finally, the forecasting performance of $\PIAR{2}{5}$ model is investigated for the out-of-sample observations from Jan 2020 to Nov 2022, using the estimated parameters given in Table \ref{tab:ParamTablepi2ar5}. Note that the forecast result has been transformed to the original scale, and the result is provided in Figure \ref{Fig:ForecastPI2AR5}. The bottom left corner of Figure \ref{Fig:ForecastPI2AR5} specifies the result of forecast values (red line) and out-of-sample data (black line) from Jan 2020 to Nov 2022, along with the confidence intervals (blue ribbon) of the forecasts, which directly shows that the forecast result of $\PIAR{2}{5}$ is reliable. In addition, the bottom right corner of Figure \ref{Fig:ForecastPI2AR5} gives the ACF plot of the forecast errors, where we observe that the forecast errors are uncorrelated with each other and therefore, the $\PIAR{2}{5}$ model effectively captures the randomness of the series. Overall, we are satisfied with the out-of-sample forecasting performance of the $\PIAR{2}{5}$ model.

\begin{figure}[ht]
	\centering
	\begin{subfigure}[b]{0.45\textwidth}
		\centering
		\includegraphics[width=\linewidth]{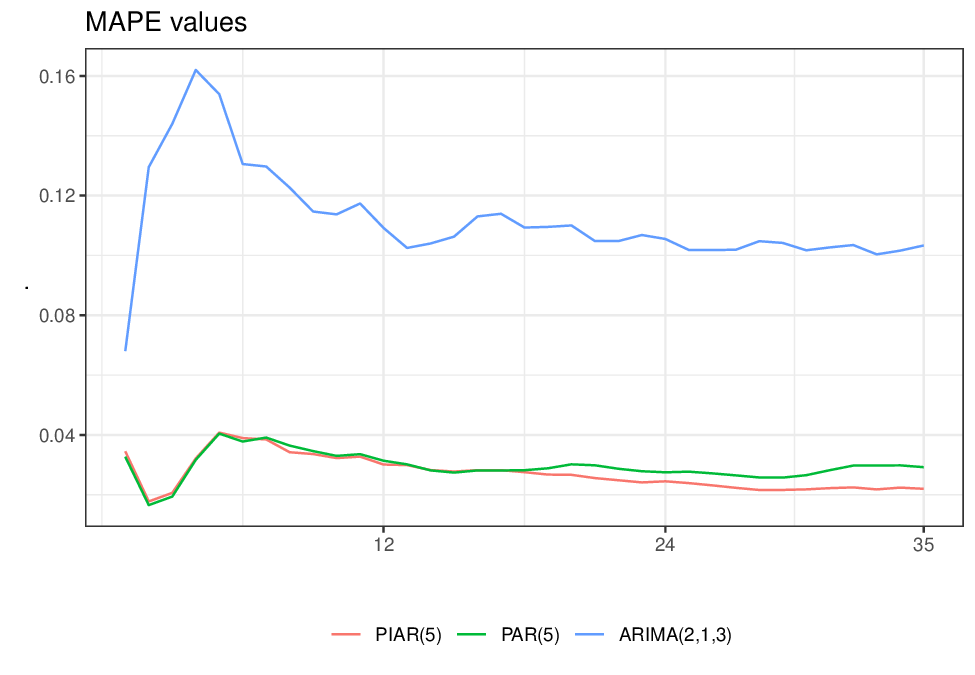}
		\caption{Mean absolute percentage forecast error (MAPE)}
		\label{Fig:MAPEvalue}
	\end{subfigure}
	\begin{subfigure}[b]{0.45\textwidth}
		\centering
		\includegraphics[width=\linewidth]{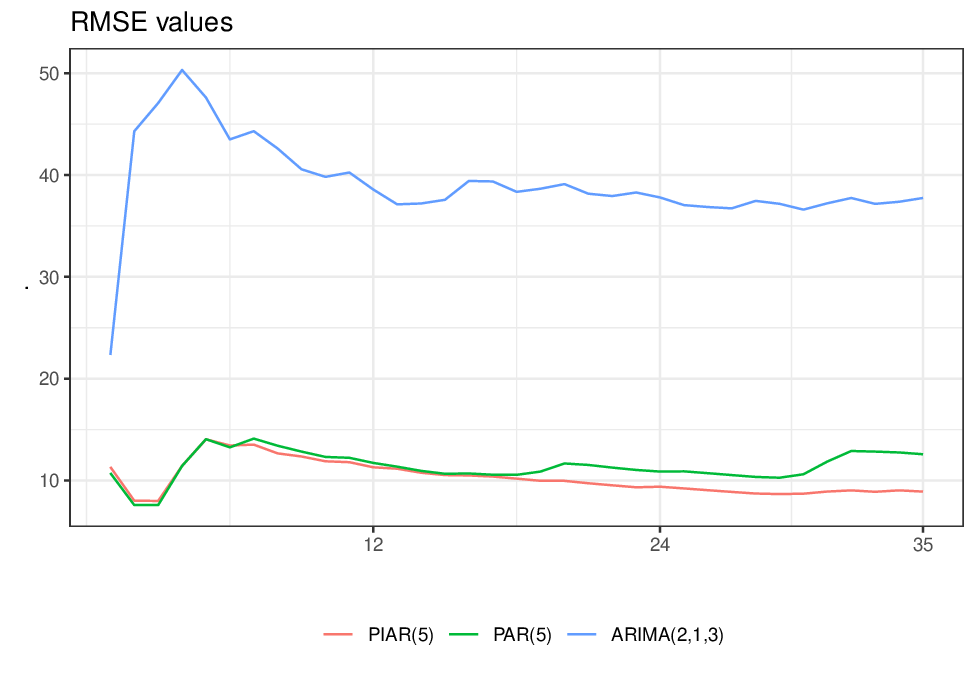}
		\caption{Root mean squared forecast error (RMSE)}
		\label{Fig:RMSEvalue}
	\end{subfigure}
	\caption{Forecasting performance of $\PIAR{2}{5}$ (depicted by the red line), PAR(5) (depicted by the green line) and ARIMA(2,1,3) (depicted by the blue line)  based on MAPE (left) and RMSE values (right).}
	\label{Fig:MAPEandRMSE}
\end{figure}

In comparison, we also use ARIMA(2,1,3) and PAR(5) models to produce the out-of-sample forecasts for our monthly electricity end use data. Note that the order of ARIMA model is automatically determined by using `forecast' package in R \citep[see][]{HyndmanForecast}. Figure \ref{Fig:MAPEandRMSE} provides the forecast performance of $\PIAR{2}{5}$, PAR(5) and ARIMA(2,1,3) models in terms of the MAPE and RMSE values. It is obvious to see both the MAPE and RMSE values of ARIMA(2,1,3) model are significantly higher compared to the values of PAR(5) and $\PIAR{2}{5}$, which suggests there is a noticeable increase in forecast accuracy by using periodic models. Moreover, the forecasting performance of periodically integrated autoregressive model $\PIAR{2}{5}$ seems to be more accurate than periodic autoregressive model PAR(5) when considering a longer forecast horizon. Therefore, we choose $\PIAR{2}{5}$ as the final model to produce forecast values for U.S. monthly electricity use.

\section{Conclusion}
\label{Conclusion}

In this paper, we have introduced and applied the multi-companion method for the analysis of PIAR models. This innovative approach relies on the eigen information of the multi-companion matrix, when expressing the PIAR models in their multi-companion representations. We find that by representing the multi-companion matrix into its Jordan canonical form, both the similarity and the Jordan matrices play important roles. The properties of the Jordan matrix are employed to propose a general definition of periodic integration, which extends the existing body of literature \citetext{see \citealp{osborn1988seasonality} and \citealp{boswijk1996unit} for example} beyond its previous focus exclusively on quarterly periodic integration of order one. 

Moreover, given that the PI-parameters can be parametrized in terms of the seed-parameters of the multi-companion matrix, we propose a new estimation approach which departs from the conventional method of directly estimating PI-parameters. This new approach initiates the estimation process by first determining the seed-parameters, which then serve as a bridge to derive the estimation of PI-parameters based on parametrization results. As a result, this approach offers a significant advantage over the existing methods \citep[see][]{boswijk1997multiple, lopez2005periodic}, which require dealing with the non-linear restrictions between PI-parameters. Additionally, our method expands the scope of analysis from the estimation of quarterly PIAR models to more general cases.

On the other hand, to validate and demonstrate the robustness and effectiveness of our multi-companion method for the estimation and forecasting of PIAR models, we have conducted both a simulation study and a practical application.  

The results of this paper offer valuable insights for the analysis of periodically integrated series. These insights can be employed to explore various aspects, including the identification of common stochastic trends, the investigation of cointegration and periodic cointegration in macroeconomic series. Moreover, given the prominence of unit root tests in non-periodic time series analysis,  it is worthwhile to advance the research in the domain of unit root tests for periodically integrated series. Our multi-companion method, as demonstrated in this paper, holds promise for further exploration and application in this context.


\bibliographystyle{unsrtnat}
\bibliography{references}  






\end{document}